\documentclass[format=acmsmall, review=false, screen=true]{acmart}
\usepackage{soul}
\usepackage{booktabs} 

\usepackage[ruled]{algorithm2e} 

\SetAlFnt{\small}
\SetAlCapFnt{\small}
\SetAlCapNameFnt{\small}
\SetAlCapHSkip{0pt}
\IncMargin{-\parindent}

\acmJournal{PACMHCI}
\acmVolume{2}
\acmNumber{CSCW}
\acmArticle{145}
\acmYear{2018}
\acmMonth{11}
\copyrightyear{2018}

\setcopyright{acmlicensed}

\acmPrice{15.00}\acmDOI{10.1145/3274414}

\begin{document}
\title[Socio-spatial Self-organizing Maps]{Socio-spatial Self-organizing Maps: Using Social Media to Assess Relevant Geographies for Exposure to Social Processes}

\author{Kunal Relia}
\orcid{}
\affiliation{%
  \institution{New York University}
  \streetaddress{Tandon School of Engineering}
  \country{USA}}
\email{krelia@nyu.edu}

\author{Mohammad Akbari}
\orcid{}
\affiliation{%
  \institution{New York University}
  \streetaddress{Tandon School of Engineering, Center for Data Science}
  \country{USA}}
\email{akbari@nyu.edu}

\author{Dustin Duncan}
\orcid{}
\affiliation{%
  \institution{New York University School of Medicine}
  \streetaddress{Department of Population Health}
  \country{USA}}
\email{dustin.duncan@nyumc.org}

\author{Rumi Chunara}
\orcid{}
\affiliation{%
  \institution{New York University}
  \streetaddress{Tandon School of Engineering, Center for Data Science, College of Global Public Health}
  \country{USA}}
\email{rumi.chunara@nyu.edu}

\begin{abstract}
Social media offers a unique window into attitudes like racism and homophobia, exposure to which are  important, hard to measure and understudied social determinants of health. However, individual geo-located observations from social media are noisy and geographically inconsistent. Existing areas by which exposures are measured, like Zip codes, average over irrelevant administratively-defined boundaries. Hence, in order to enable studies of online social environmental measures like attitudes on social media and their possible relationship to health outcomes, first there is a need for a method to define the collective, underlying degree of social media attitudes by region. To address this, we create the Socio-spatial-Self organizing map, ``SS-SOM'' pipeline to best identify regions by their latent social attitude from Twitter posts. SS-SOMs use neural embedding for text-classification, and augment traditional SOMs to generate a controlled number of non-overlapping, topologically-constrained and topically-similar clusters. We find that not only are SS-SOMs robust to missing data, the exposure of a cohort of men who are susceptible to multiple racism and homophobia-linked health outcomes, changes by up to 42\% using SS-SOM measures as compared to using Zip code-based measures.\end{abstract}

%
%


%
%


\maketitle

\setlength{\abovedisplayskip}{1pt}
\setlength{\belowdisplayskip}{1pt}

\renewcommand{\shortauthors}{K. Relia et al.}

\section{Introduction}
Public health, economic and other social outcomes depend on what we experience around us. Thus there is a need to identify the relevant environment around an individual in relation to a specific exposure. For example, in which areas of a city are we likely to feel happiness? Or conversely, where will we experience discrimination? In particular, many health outcomes depend on environmental exposures; it has been shown that there is a link between social exposures like racism and homophobia to disease and risky health behaviors \cite{jee2015black, frye2010neighborhood,carpiano2011community}. There are many hypothesized mechanisms for this link. Discrimination against black workers in employment can lead to lower income and greater financial strain, which in turn have been linked to poor health outcomes \cite{darity2003employment}. Racism may also directly impact health by engaging psycho-biological mechanisms induced in the stress response \cite{clark1999racism}. A major challenge to this work has been obtaining measures of discrimination in the community. It has been shown that it is difficult to measure sensitive attitudes like racism or homophobia via surveys, due to biases like social desirability bias \cite{tourangeau1991measuring, mackenzie2012common}. Survey-based measures are costly and laborious to obtain widely or with high granularity, such as at within-city levels. Moreover, mapping results to Zip codes results in unnecessary and large spatial averaging \cite{duncan2013examination}. Such attitudes can also be dynamic (e.g. socio-political events could cause increases in discrimination), and deploying surveys at an appropriate timescale to discern these changes would be labor and cost intensive. Measures of discrimination such as from hate crime reports also provide an incomplete picture of discrimination; they only capture discrimination that results in a crime being reported. As well, hate crime reports can also suffer from biases affecting the type of reports and/or who contributes them. 

In social computing research, social media has been shown to provide a unique window into the social experience of people, and in particular Twitter has been used for assessing sensitive topics, such as discrimination \cite{davidson2017automated,silva2016analyzing,Chaudhry2015,de2016social}. In order to use such measures of discrimination in downstream analyses, such as studying the link between online discrimination reports and health, we need a way to assess the latent sentiment in a location from the Tweets. This is challenging because a single Tweet may be noisy (e.g. its content regarding racism could be unclear). As well, when resolved to very high-resolution geographic areas, Twitter data can be sparse. Further, a single Tweet may not be representative of the social sentiment in an underlying area. Therefore, there is a need for methods to define the collective latent sentiment of an area based on inconsistently distributed point-observations from geo-located data such as Tweets. 


Accordingly, we present a social computing pipeline for identifying relevant geographical areas of social processes, as defined by text from Twitter posts. Our approach, SS-SOMs divides the city into subareas, defining the optimal collective sentiment by location. Our work accounts for the observational nature of social media by accounting for varying Tweeting levels by location. We compare the method to appropriate baselines using prevalence of homophobia (expression of a homophobic sentiment, or of an experience of homophobia) and racism (conveys a racist message, or indicates someone experiencing racism) on Twitter, as significant and understudied forms of discrimination, and conducive to being measured via social media (though the pipeline can be applied to other social processes, especially those uniquely represented on social media). We focus on New York City (NYC) as discrimination within urban cities is highly varied due to proximity of neighborhoods with dramatically different norms and cultures. As well, health disparities for affected populations are significant in cities. Using data from a cohort of racially/ethnically and socioeconomically diverse young gay, bisexual and other men who have sex with men (MSM), who are disparately affected by racism and have a unique footprint in NYC \cite{egan2011migration, collins2009we}, we compute how the relevant online social media measures of racism and homophobia for them in NYC measured via SS-SOMs would be different than via Zip codes. Therefore, this work addresses the need for new measures of racism and homophobia especially within urban cities. Specific contributions are:
\vspace{-0.075cm}
\begin{itemize}
    \item A pipeline to partition an area into regions of collective, consistent sentiments from social media data  (SS-SOMs).
    \item Evaluation of the quality of SS-SOM clusters compared to relevant baselines. \item Demonstration of impact by comparing online homophobia/racism exposure measures via SS-SOMs v.s. Zip codes for an MSM cohort in an area (NYC) where disparities are significant.
\end{itemize}


\section{Related Work}

\subsection{Characterizing Environments by Social Media}
Concentration of online activity has been used to define locations from Flickr photos \cite{Deng2009spatial}, and spatial statistics have been combined with such data to identify specific regions via particular photo tags \cite{Rattenbury2009placeFlickr} (but only identifying those locations captured in photos, and not partition an entire geographic area). Foursquare check-ins have been used to identify ``neighborhoods'' \cite{noulas2011empirical,cranshaw2010seeing,LeFalherSohoRome,CranshawLivehoodsProject}. This work has identified clusters of similar types of check-ins to characterize the physical environment based on activity type (e.g. shopping area versus restaurants, etc.). This line of work only focused on specific types of areas, is focused on identifying those specific ``neighborhoods'' and not partitioning an area comprehensively. Other work has assessed location data and the reliability of social media tags  \cite{hollenstein2010exploring}. Finally, sound types have been identified via social media. Sound is a localized phenomena which can be mapped to precise locations (and can be compared to other data at that level), instead of by region \cite{aiello2016chatty}. Notably, the above approaches are built on observational, numerator-only data -- thus resulting clusters can be confounded by areas where or types of locations in which individuals tend to check-in more frequently \cite{chunara2017denominator,noulas2011empirical}. This matters less when only specific locations/neighborhoods are being identified, compared to if partitions of an entire area are desired.

As the relevant environment around us goes beyond the physical and includes social  processes around us, our work is novel because it considers the text of Tweets to characterize social environments (which hasn't been done from a geographical point of view). As the social environment depends on anywhere there are people, our goal is also different from the above; we aim to partition (e.g. create contiguous clusters) to allow for understanding of exposures at any given location. Moreover, the task of describing the \emph{social} environment helps us define constraints for the partitioning method. We require a set of subareas that are collectively exhaustive for the area they divide, contiguous and mutually exclusive, and each subarea should represent an exposure level that best exemplifies all of the individual social media posts they are defined by.

\subsection{Methods for Identifying Spatial Structure and  Generating Boundaries}
Generating appropriate boundaries is an active research area, given the increasing amount of geo-located data. However the specific challenge of defining areas of consistent social attitude is different from previous work. For such social attitudes, individual Tweets can be noisy (the text of an individual Tweet may not be clear regarding the attitude), are not consistently generated everywhere, and to be useful in assessing health outcomes, must be linked to a unique (non-overlapping) area representing the underlying sentiment. Given these constraints, we are specifically charged with developing homogeneous, contiguous partitions of the specified geography (a continuous field, for example using kernel methods would not be appropriate for the application). 

A variety of methods can be applied to uncover hidden spatial structure in geographic data, including clustering \cite{Deng2009spatial}, density estimation \cite{greenTweitterAuthor,Jones2014redefiningNbhd} and neural networks \cite{galster2008quantifying}. To define location representations/boundaries from Flickr and Foursquare check-in data some approaches have harnessed burst-analysis techniques which model the distribution probabilistically, highly peaked over a small number of more nearby values \cite{Rattenbury2009placeFlickr} or common clustering methods such as DBSCAN (Density Based Spatial Clustering of Applications with Noise) which is an algorithm for noisy data \cite{intagorn2011learning, srivastava2015geo}, K-means clustering \cite{LeFalherSohoRome}, and DBSC (Density-Based Spatial Clustering) which focuses on content similarity and spatial proximity equally but doesn't guarantee to partition a region \cite{liu2012density}. 

Other work identified irregularities in amount of Tweeting by location over time \cite{lee2011discovery} using K-Means clustering and Voronoi polygons \cite{frias2012characterizing}. In epidemiology, environmental exposures are traditionally quantified via Zip codes and census tracts \cite{chunara2013assessing}. While these approaches (Voronoi polygons, Zip codes) fulfill the criteria for our social process area partitions: they define a set of subareas that are collectively exhaustive for the area they divide, and are contiguous and mutually exclusive, the resulting regions are defined administratively or based on amount of data and not in a manner relevant to the exposure. Therefore computing the average social attitude over these areas will incur unnecessary spatial averaging. 


A sophisticated approach for defining geographic areas uses artificial neural networks (ANNs); an unsupervised learning approach \cite{Kohonen1990SOM}. The input signal (vector containing information about the attributes of data to be mapped) is linked to a spatial location and the ``self-organized map'' (SOM), is organized based on the amplitude of these signals. Many different adaptations of SOMs have been proposed spanning organization of the input vector \cite{yin2002data}, algorithm \cite{deng2000esom,kiviluoto1996topology}, or layout of the output space \cite{liou1996handprinted}. There has been a specific focus on preserving information about the topological distance between input nodes. Many such modifications can be grouped into 3 approaches: (1) including geo-coordinates as a part of the input vector, (2) calculating topological distance between output nodes instead of the distance between the weight of the nodes to localize the SOMs \cite{deng2000esom} and, (3) changing the SOM ``neighborhood'' function to cover a wider width \cite{kiviluoto1996topology}. Modifications on inclusion of geo-coordinates have included: (a) using a combination of the weight vectors and neuron spatial positions to measure topological distance between points and cluster them together \cite{yin2002data,kitani2011exploring} and (b) searching for the best matching unit (nearest node) only within a predefined topological vicinity (called Geo-SOMs) \cite{baccao2004geo}. Both of these modifications lead to well-defined but overlapping clusters. Consequently, a major shortcoming of this is the difficulty in developing contiguous positions of resulting areas \cite{guo2005multivariate}. Despite these varied approaches, to our best knowledge, there is no method that guarantees the resulting clusters to be at once topologically constrained, contiguous, non-overlapping, and allowing for control over the number of clusters formed.

\subsection{Data for Tracking Racism/Homophobia}
Geographers and social scientists have examined sentiments like racism for decades. 
Causal mechanisms between racism and health outcomes have been clearly established \cite{jee2015black}. These conclusions have been reached in multiple settings, like in the workplace, racial discrimination has been shown to relate to adverse health outcomes \cite{darity2003employment} -- the link being slightly larger for men versus women, but significant for both. More broadly are community-level influences, which are why here we focus on social media data across and within a city. Research has shown that neighborhood influences like racism and homophobia can play an important role in influencing health behaviors including substance use and condomless sex among MSM \cite{frye2010neighborhood,carpiano2011community}. Other recent work has shown links between racism in communities and mental health outcomes specifically in black populations \cite{bor2018police}. 

One approach to assess racism has been based on geographic clustering or concentration of different racial groups, to extract measures of segregation \cite{White1983spatial}. While segregation is an important factor in ascertaining racism, it doesn't directly speak to the experience of individuals. Approaches to measure experience with racism have largely been through survey mechanisms \cite{Aosved2009intolerence}. Concerns about how well surveys capture all aspects of the prejudice have been highlighted (e.g. based on terms being stigmatized, social censoring, different population groups having different self-report biases, and the validity and reliability of measures) \cite{Costa2013systematic}. A new proxy for an area's racial animus from a non-survey source is the percent of Google search queries that include racially charged language \cite{stephens2014cost}. Here it was noted that Internet-reported data is not likely to suffer from major social censoring, and is a medium through which it is easier to express socially taboo thoughts \cite{Kreuter2008desirability}. More recently, Twitter has been used to identify and spatially map racism \cite{Chaudhry2015}. This work has parsed Twitter data via keyword filtering, summarized and compared example Tweets expressing racism. 

More recently, Twitter has also been used to systematically assess the main targets of online hate speech (defined as bias against an aspect of a group of people, thus broader than just racism/homophobia in online social media) \cite{silva2016analyzing}.  Notably from this work, race and sexual orientation were amongst the top hate speech targets on Twitter (as well as Whisper). Overall, the prevalence of hate speech on social media is considered a serious problem. Finally, because we aim to understand racism and homophobia as comprehensively as possible spatially, we include both reports of exposures and racist/homophobic statements, both of which are possible on Twitter. 



\section{Data}
\subsection{Twitter Sample} \label{section:trainingdata}
The Twitter Application Programming Interface (API) was used to source geo-located Tweets having point coordinates within the boundaries of New York City (NYC). We initially used a generous bounding box enclosing all five NYC boroughs: 40.915256 N to 40.496044 N and 73.700272 W to 74.255735 W  \cite{nagar2014case}. Tweets that were outside the precise NYC boundaries were later filtered out while mapping the Tweets to grid cells, as discussed in  section \ref{section:TweetingLevelsoverSpace}. The time period for Tweets was selected to exactly overlap the time period of the MSM cohort mobility data (January 25, 2017 to November 3, 2017). This resulted in 6,234,765 Tweets. Data from NYC was used as it provides an ideal environment based on data volume, population density, and also addresses the increasing imperative to understand social and health disparities specifically within urban areas \cite{halkitis2013sociodemographic}. 

\subsubsection{Training Data for Racism Classification}
In line with the typical size of training data sets for Twitter classification efforts, we used 9785 Tweets for the training set \cite{waseem2016hateful}. As the topic is very nuanced, to capture a comprehensive set of training examples we used a combination of keyword filtering, labelling, and iterative learning to generate this training data.

\begin{table}[b]
\caption{Top Twitter features after classification.}
\label{tab:topfeatures}
\begin{minipage}{\columnwidth}
\begin{center}
\begin{tabular}{l| l| l}
  \toprule
  {\small \textbf{\begin{tabular}[c]{@{}l@{}}Classifier\\ \end{tabular}}}
   & {\small \textbf{\begin{tabular}[c]{@{}l@{}}Racism top features\\ \end{tabular}}}
   & {\small \textbf{\begin{tabular}[c]{@{}l@{}}Homophobia top features\\ \end{tabular}}}\\
  \hline
  Neural & useless\_nigger, white\_trash, coon, & hate\_same\_sex, gay\_marriage,\\
  Model & stop\_chinaman, racist\_bigot & son\_is\_gay, hooker, lads\_kissing\\
  \hline
  SVM & me\_a\_nigger, ching\_chong\_chinaman, & gay, gay\_people, son, I\_hate, \\
  & white\_trash, suck\_it\_niggers, useless\_beaner & being\_gay\_is\\
  \bottomrule
\end{tabular}
\end{center}
\bigskip\centering
\end{minipage}
\end{table}%
\vspace{-0.25cm}
\paragraph{Keywords for Racism Classification}
We used findings from the existing literature on social media and various forms of discrimination \cite{stephens2014cost,Chae2015blackMortality,Chaudhry2015}, in combination with Urban Dictionary and Wikipedia to generate a dictionary of keywords (Appendix) to source ``racism'' Tweets. As discussed above, in order to ensure we were capturing Tweets that represented exposure to racism comprehensively, our definition for a ``racism'' Tweet is one that conveys a racist message, or indicates someone experiencing racism. We used this dictionary to source 100 recent Tweets for each keyword from the Twitter API. This initial keyword filter resulted in 3868 Tweets. Notably, some keywords were derived from a previously developed Internet-based measure of ``area racism'' that did not rely on the provision of responses to survey questions and thus is less susceptible to social desirability bias \cite{stephens2014cost}. Further, we highlight that deciding what is a ``racism'' Tweet is a nuanced issue and we wanted to make sure not only that the keywords selected were consistent with other recent papers using Internet-data to measure racism, but that the data did not over-represent colloquial language. For example, not all text that contains the ``n-word'' are motivated by racist attitudes. Specifically, more colloquial forms of the ``n-word'' (i.e., ending in ``-a'' or ``-as'' vs.``-er'' or ``-ers'') were not included given that prior work found that these versions were used in different contexts \cite{stephens2014cost}. Stephens et al. found that after this exclusion, prevalence of the resulting terms did relate to racial attitudes in a geographic area measured through disparate voting practices for a black candidate \cite{stephens2014cost}. To further ensure the keywords selected do not result in colloquial language being included, we also examined the top features post-classification (using the final training data) (Table \ref{tab:topfeatures}) which confirmed the resulting labels were indicative of our criteria and not colloquial language. 
\vspace{-0.25cm}
\paragraph{Labelling of Tweets by Human Annotators}





We clearly defined the criteria for labelling a Tweet as indicating ``racism'' (versus ``no racism'') as described above, to Amazon Mechanical Turk (AMT) workers for annotation, and through initial trial experiments confirmed the clarity of our instructions (included in the Appendix). As this task involved exposure of humans to potentially sensitive content, we clearly indicated the task is about racist Tweets, and created each Tweet as an individual Human Intelligence Task (HIT) giving workers a chance to discontinue at any point without losing payment if they felt uncomfortable.
The workers were paid the minimum rate suggested by AMT when allocating the task (resulting in the amount paid being equivalent to \$14.40/hr \cite{finin2010annotating}). Each Tweet (only the text of the Tweet was provided) was labelled by two workers on AMT and Cohen's kappa statistic was 0.62 \cite{cohen1960coefficient}. Our team member manually examined Tweets with a disagreement to resolve the label. In the end, 2040 of the 3868 Tweets sourced by keyword filtering were labelled as expressing ``racism''. To add data from the negative class to our training set, in a ratio similar to other work, we added 4188 Tweets without any racist keywords in them \cite{waseem2016hateful,burnap2015cyber}.

\vspace{-0.25cm}
\paragraph{Iterative learning}
Finally, to improve our training data set even further we aimed to find and include labels for Tweets that may be difficult to classify. To do this, we used an iterative learning approach as previously implemented with Twitter data \cite{liu2017assessing, huang2017high} to ensure our classifier was clearly capturing the definitions of racism and homophobia we stated. In this step we classified approximately 10000 unlabelled Twitter posts (from the NYC set used for this project), using the 8056 labelled Tweets from the previous section. We used the SVM classifier as described in section \ref{section:SVM}. We then manually labelled the edge cases (approximately 5\% Tweets with an output label from classification that was near the threshold). We added these labelled edge cases to our training data set with the aim of increasing the number of edge labels in the training data which may improve the classifier's performance \cite{huang2017high,liu2017assessing}. We did this successively until the SVM classifier performance plateaued (after three iterations). In the end, this increased the training data size from the 8056 to 9785 Tweets, 2398 of which indicated ``racism'' and 7387 ``no racism'' \cite{waseem2016hateful}.


\subsubsection{Training Data for Homophobia Classification} 
The same procedure as above (keyword filtering, labelling by human annotators and iterative learning) was used to create the training data for Tweets that express or represent an experience of homophobia. 
\vspace{-0.25cm}
\paragraph{Keywords for Homophobia Classification}
In contrast to the study of racism on social media, there is not as much precedence in the literature regarding social media expressions of homophobia. Therefore the set of training Tweets for homophobia derived using keywords from Urban Dictionary and Wikipedia alone was not as comprehensive as the resulting Tweets for racism. Thus we added Tweets from the ``@homophobes'' Twitter account. The explicit purpose of this account is to Re-Tweet homophobic Tweets from anywhere, so this would increase the linguistic diversity of the training data set. All 1626 Tweets and Retweets as of December 2017 were obtained from the account ``@homophobes'' via the Twitter API, and added to 1177 recent Tweets containing the homophobia keywords listed in the Appendix. 
\vspace{-0.25cm}
\paragraph{Labelling of Tweets by  Human Annotators}
We clearly defined the criterion to annotators that a Tweet labelled as ``homophobia'' (versus ``no homophobia'') should either express a homophobic sentiment, or express an experience of homophobia. Tweets (only the text of the Tweet was provided) were labelled by two workers on AMT and Cohen's kappa statistic was 0.71. Disagreements were resolved by our team. At the end of the labelling stage, 1990 of the 2803 Tweets sourced in the previous section were labelled as ``homophobic''.
\vspace{-0.25cm}
\paragraph{Iterative learning}
Following the same approach as in racism classification, SVM and iterative labeling were used to classify the homophobia Tweets. Finally, this resulted in a training data set of 2176 ``homophobia'' Tweets, and 7535 ``no homophobia'' Tweets.

\subsection{MSM Cohort Data} To demonstrate the effect of measuring online social discrimination through SS-SOMs versus other spatial partitioning methods, we use mobility data from a cohort of young MSM who are part of the National Institutes of Health funded P18 Cohort Neighborhood Study \cite{halkitis2013sociodemographic}. It is important to calculate the potential exposure-space for this group specifically, as 1) they are susceptible to multiple racism and homophobia-linked health outcomes, and 2) the locations they frequent and thus their environmental risk-space is unique compared to other populations \cite{collins2009we, koblin2013methods, egan2011migration, tobin2013social}. P18 is an ongoing prospective cohort study of sexual behavior, substance use, and mental health burdens that includes racially/ethnically and socioeconomically diverse young MSM. We measure the exposure of racism and homophobia specifically on this cohort of MSM because it is widely known that experiences of social discrimination among the members of minority groups like MSM can lead to poor health outcomes \cite{choi2011strategies}.

Participants were eligible for the original study if they were 18 or 19 years old, biologically male, lived in the NYC area, reported having had sex with another male in the 6 months before screening, and self-reported a HIV-negative or unknown serostatus. Those in the neighborhood part of the study consent to carry a GPS tracker for two weeks. The tracker logs the precise geo-location every 10 seconds. In total, at time of this paper preparation, we had mobility data from 226 men. Data was collected between January 25 and November 3, 2017. The minimum amount of time an individual wore the tracker was 6 days, maximum was 82 days. The mean duration was 21.3 days.


\section{Socio-spatial Self-organizing Map pipeline}
\begin{figure}
  \includegraphics[trim=0.9cm 8.5cm 1.3cm 2cm, clip, scale=0.6]{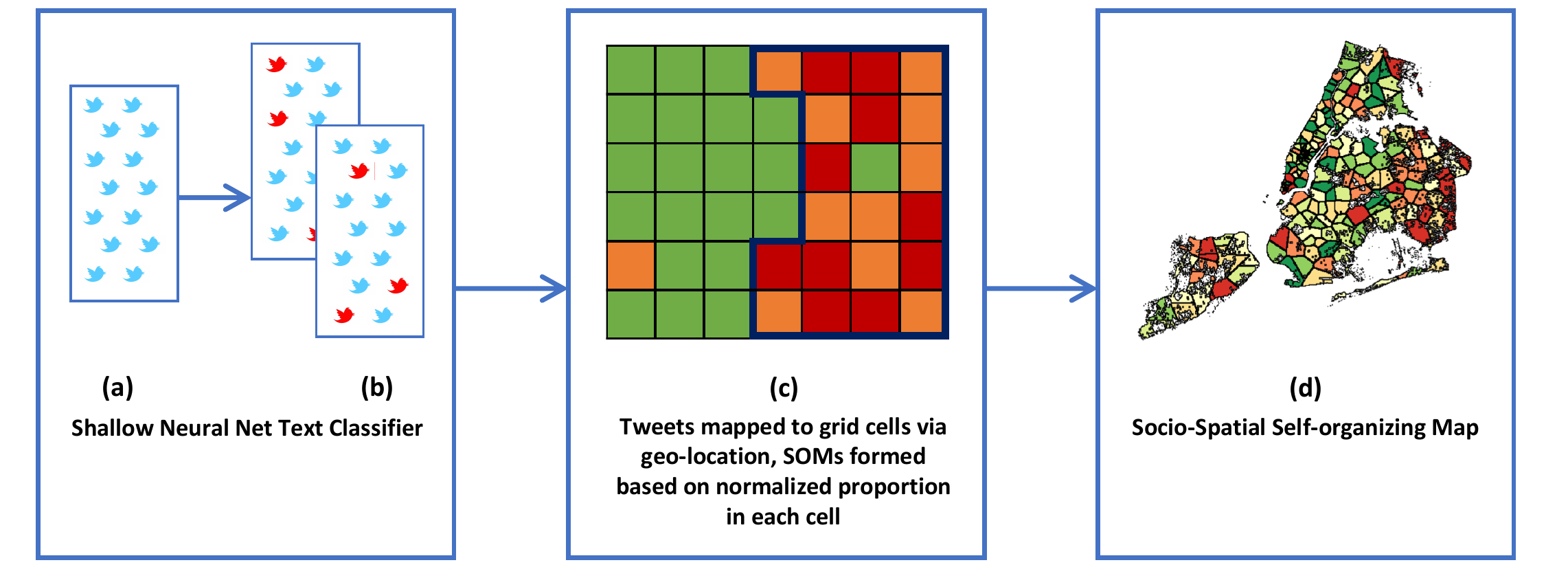}
  \caption{The SS-SOM pipeline. (a) Tweets made within the NYC bounding box are classified as (b) racism vs. no racism and homophobia vs. no homophobia using a shallow neural network, where the resulting probabilities of Tweet is used to classify the Tweet. (c) NYC is divided into grid cells, geo-location of each Tweet is used to map Tweets to the cells, and the normalized count is represented by different colors of the grid cells - from red (high racism/homophobia) to green (no racism/homophobia). SS-SOM clustering is performed (example boundary represented by the dark blue border). (d) Resulting SS-SOM map for prevalence of racism in NYC.}
  \label{fig:pipeline}
\end{figure}

The SS-SOM pipeline will partition an area into sub-areas that are collectively exhaustive, contiguous and mutually exclusive, where each subarea represents homogeneous exposure that best exemplifies the individual social media posts they are defined by. Implementation of this method is a three staged process where (1) Tweets are classified using a neural network and the training data generated above, (2) classified Tweets are then mapped to grid cells used to divide the entire NYC to account for varying level of Tweeting patterns across the city and (3) an augmented version of the Self-organizing Map is used to partition the city into the regions that are most similar in the social attitude, and contiguous. Figure \ref{fig:pipeline} illustrates these steps.

\subsection{Tweet Classification using Neural Models}
To classify Tweets we use neural learning models, which have been shown to have improved performance over traditional classifiers \cite{tang2014learning,dos2014deep}, especially for short texts such as Twitter messages which contain limited contextual information. As the models combine the small text content with prior knowledge, they therefore perform better than models using just a bag-of-words approach. To implement the model, we create an embedding for words in our Twitter data set and use cosine distance to measure similarity between those Tweets in the embedded space. This approach is similar to the popular Word2vec model \cite{goldberg2014word2vec}, but is trained on our actively learned racism and homophobia training data sets (section \ref{section:trainingdata}). The model is a shallow, two-layer neural network that is trained to reconstruct linguistic context of words from its training text corpus input. The model produces a high-dimensional vector space, with each unique word in the corpus being represented by a vector in the space. Thus words that share common contexts in the training corpus are located in close proximity to one another in the space. We validate this model by comparing its performance to a baseline classifier (Support Vector Machines (SVM)) \cite{JoachimsTextCategorization}.

\subsubsection{Shallow Neural Network}
In our model we maintain an $n$-gram approach coherent with the baseline SVM classifier (section \ref{section:SVM}); the features used for the embedding ranged from one to six words, half the number of average words per Tweet \cite{greenTweitterAuthor} (character n-grams did not improve performance as in other work with similar size data sets \cite{huang2017high}). Given a word vocabulary $W$ of size $N$ derived from the labeled training data, $w_i$ : $i$ $\epsilon$ $\{1, ..., N\}$, where the goal of neural embedding is to learn a distributed representation for each word $w_i$ accounting for the semantic similarity of words while simultaneously computing the probability of distribution over the predefined classes. More specifically, the objective of embedding is to minimize the negative log-likelihood: $- \frac{1}{N}\sum\limits_{t=1}^{N} y_i log(f(x_i))$ where $x_i$ is the $n$-th document, $y_i$ is the label, and $f$ is an embedding function which maps the text document into a vector representation. Following research efforts in word embedding, we first embed each word in the text with an embedding vector and then employ the average of word embeddings to form the tweet embedding. This is in turn fed to the final layer, a softmax function, to define the probability of a class given the embedding of the input text. We will then have two outputs for each Tweet: the probability of being racist, and non-racist (and similar for homophobia), Tweets classified using a probability threshold of 0.5.



\subsection{Adjusting for Varying Spatio-Temporal Tweeting Levels}

\subsubsection{Tweeting Levels over Space}\label{section:TweetingLevelsoverSpace}
In order to account for varying levels of Tweeting by location, we normalized the number of racism/homophobia Tweets by the total number of Tweets by dividing NYC into a series of grids (spatial bins), and using these grids with a normalized proportion in subsequent classification analyses. Work using Foursquare data has shown that the geographic size of the bins is important to consider as it can alter the resulting spatial distribution \cite{preoctiuc2013exploring}. Here we discuss how we chose an appropriate spatial bin size that would account for the spatially inconsistent nature of Twitter data, but also minimize averaging and allow us to harness the precise geo-location information associated with Tweets. 

To divide the city into grids we used available latitude/longitude coordinates of the sourced Tweets (which have 8 decimal places). By systematically rounding off each latitude/longitude pair to a lower precision (7 to 1 decimal place), we divided the city into grid cells at each resolution. For example, a Tweet with co-ordinates 40.83470082,-73.92287411 is rounded off from  (40.8347008,-73.9228741) to (40.8,-73.9). Based on these coordinates, the example Tweet is mapped to the corresponding grid cell. Table \ref{tab:NYCgridcount} shows the results by precision level.

From the grid count and average Tweet count per grid (Table \ref{tab:NYCgridcount}), we selected 3 decimal point precision, as this offered a balance between a substantial number of grids (using more grid points significantly increased computation time for analyses), while still representing a reasonably local geographic area and high enough proportion of racism/homophobic Tweets per grid cell. Later (in section \ref{section:SS-SOM}), we show that this also resulted in a close relation between the number of SS-SOMs and Zip codes, which was important so that any findings related to cluster consistency are related to the actual clusters and not increased spatial averaging due to differences in the average size/number of clusters. We then removed grid cells that were not within the strict NYC boundaries via borough boundary shape files\footnote{ https://data.cityofnewyork.us/City-Government/Borough-Boundaries/tqmj-j8zm/data}, leaving 72,484 grids.

We highlight that while the normalized values by grid account for an unequal distribution of Tweets by location, it would be insufficient to use these grids alone to assess the prevalence of racism or homophobia by location due to an unequal distribution of Tweets per grid cell. Further, we note that we also tested a user-centric versus Tweet-centric approach to both grids and clusters, to assess if ``loud'' users may be skewing results. The results showed that there could be some loud users, but they are not changing results based on content. Pearson's correlation was 0.87 for racism and 0.91 for homophobia, between user-centric and Tweet-centric formation of grids. Cluster similarity using user-centric and Tweet-centric binning was 0.93 and 0.95 for racism and homophobia respectively. Thus all further analyses use a Tweet-centric approach in order to capture the impact of all of the \emph{content}.

\begin{table}%
\caption{Result of dividing NYC bounding box into grids and rounding coordinates to selected precision. The area covered by each grid cell for 3 decimal places is 9364 m$^2$, which is similar to the area of 1 block in NYC (1 block is the area enclosed between 2 streets and 2 avenues $\approx$ 11,100 m$^2$).}
\label{tab:NYCgridcount}
\begin{minipage}{\columnwidth}
\begin{center}
\begin{tabular}{l| l| l| l| l| l}
  \toprule
  {\small\textbf{\begin{tabular}[c]{@{}l@{}} Decimal\\ Places\end{tabular}}}
    & {\small \textbf{\begin{tabular}[c]{@{}l@{}}Number of\\ Grids\end{tabular}}}
    & {\small \textbf{\begin{tabular}[c]{@{}l@{}}Area of\\ Each Grid Cell\end{tabular}}}
    & \multicolumn{3}{c}{\small \textbf{\begin{tabular}[c]{@{}l@{}}Avg. Tweets per  Grid\end{tabular}}}
    
    \\  \cline{4-6}

    {}&{}& {(in $m^2$)}
    &{\small \textbf{Total}}
    & {\small \textbf{Racism}}
    & {\small \textbf{Homo.}}
       \\
    
    \hline
    1 & 35 & 9.367x$10^{7}$ & 178,136.143& 1738.629 & 3135.200\\
    2 & 2,206 & 9.365x$10^{5}$ & 2,826.276& 27.585 & 49.743\\
    \textbf{3} & \textbf{150,385} & \textbf{9.364x$\mathbf{10^{3}}$} & \textbf{41.459}& \textbf{0.405} & \textbf{0.730}\\
    4 & 2,796,484 & 9.364x$10^{1}$ & 2.230& 0.022 & 0.039\\
    5 & 10,620,968 & 9.364x$10^{-1}$ & 0.587& 0.006 & 0.010\\
    6 & 14,466,241 & 9.364x$10^{-3}$ & 0.431& 0.004 & 0.008\\
    7 & 14,768,312 & 9.364x$10^{-5}$ & 0.422& 0.004 & 0.007\\
  \bottomrule
\end{tabular}
\end{center}
\bigskip\centering
\end{minipage}
\end{table}%

\subsubsection{Tweeting Levels over Time}
As we controlled for the Tweeting levels over space by dividing the city into grid, we also assessed if Tweeting levels change over time and if that may influence results. We assessed the proportion of homophobia or racism Tweets each month using student's t-test to assess if there was any significant difference between each two adjoining months' normalized homophobia and racism values. Results of the tests showed no significant difference (p-values ranged from 0.54-0.91) between any adjacent months, indicating no major underlying differences by month. Hence, classification results from each of the included months are similar and results would not differ by time.

\subsection{Socio-Spatial Self-Organizing Map (SS-SOM)}\label{section:SS-SOM}
Self-organizing Maps (SOMs) are frequently used in geographical clustering, the main difference being that the constraint of contiguous clusters is not imposed. The broad approach for implementing any version of SOMs remains fairly similar. Details of the SOM algorithm can be found elsewhere \cite{Kohonen1990SOM,intelligence2012introduction}, and here we focus on the specifics and unique aspects of SS-SOMs. Implementation can broadly be divided into 2 stages: (1) initialization, and (2) organization.
\begin{itemize}
    \item \underline{Initialization}: Similar to all SOMs, SS-SOM require an input vector (of weights, which in our case is the proportion of Tweets), output vector (initially consists of just the geographic locations of the nodes, and eventually to have linked information about the weights), and multiple operational parameters. Operational parameters are: learning rate (rate at which winning nodes, the final centers of each cluster, approach their destination cluster), number of learning cycles, and neighborhood radius which describes how ``far away'' in terms of weight, adjoining grid cells are to be considered in each learning cycle and in our case is adapted to include geographic distance also. The modification of constraining the neighborhood radius is operationalized by specifying the distance at which nodes are considered in re-weighting of the output nodes, to ensure geographical adjacency.
    \item \underline{Organization}: Organization of the SS-SOM is similar to regular SOM except the neighborhood radius is augmented to include surrounding nodes that have similar weights in terms of both Tweet proportions (semantic similarity) and topological distance (geographic similarity).
\end{itemize}



\paragraph{Detailed description of SS-SOM} \label{section:SS-SOM_desc}
We define the following input variables for the SS-SOM algorithm: $g$, the number of grid cells, and $\mathbf{I}$, an input vector space, where length of $\mathbf{I}$ = $g$. 
The operational parameters help us specify how the SS-SOM algorithm functions: $t_{max}$ is the number of learning cycles (50), $\eta_0$ is the learning rate at time $0$, $\eta_0$ $\approx$ 0.1, $\eta_0$ $\epsilon$ $Q^+$ and $\eta(t)$ is the learning rate at time $t$: $\eta(t)$ = $\eta_0 . e^{-\frac{t}{t_{max}}}$, $\eta(t)$ $\epsilon$ $Q^+$. The operational parameters are set in line with previous work \cite{intelligence2012introduction}. Additionally, we performed a sensitivity analysis for the number of learning cycles, and found that the learning rate plateaus as we near 50 cycles (the chosen value), after which the number of resulting clusters also plateaus. Later, we found that this resulted in a number of clusters that was most comparable to the number of Zip codes at the chosen threshold value $\tau$ (Figure \ref{fig:similarity}(b)). And, since the average resulting area of SS-SOM's with this threshold and number of learning cycles were 8.39 x $10^6$ m$^2$ for racism and 7.73 x $10^6$ m$^2$ for homophobia, which are very similar compared to 3.69 x $10^6$ for Zip codes, we deemed these settings appropriate. Moreover, the average number of Tweets in each SS-SOM for racism was 647.36 (0.98\%) and for homophobia was 1075.8 (1.76\%), which was very similar to Zip codes (284.36 (0.98\%) and 512.77 (1.76\%) respectively). Hence, resulting comparisons would only be due to homogeneity within the clusters and not because of a changing number or changing area of clusters.

Each output node has two attributes, fixed position and weight. While the fixed position is a constant attribute that helps in determining the initial position of the node in the grid system (NYC map), the weight of the node is a varying attribute and determines how the clustering takes place based on the total and racism/homophobia Tweet counts. $S$ is a set of output nodes to be created: $|S|$ = $g$, $n_{i}$ is a node from the set of output nodes: $n_{i}$ $\epsilon$ $S$. Then, $\forall$ $n_{i}$, $p_{i}$ is the position of node $n_{i}$ (the coordinates of the node (grid cell's) location on the map), $p_i$ $\epsilon$ (longitude, latitude). And, $w_{i}$ = the weight of node $n_{i}$, is initialized using random values of any range, dim$(w_{i})$ = dim$(\mathbf{I})$.

We modify the traditional SOM (discussed in section  \ref{section:tradSOM}) to ensure we create topologically-constrained and non-overlapping clusters by defining the input vector space $\mathbf{I}$ with two vectors: number of Tweets in a grid cell, and number of racism/homophobia Tweets. Hence, dim($\mathbf{I}$) = 2, and correspondingly, dim($w_i$) = 2. We also define the following operational parameters specifically for the modified version of the SOM: $\phi$ is a cluster ID assigned to each grid cell that uniquely identifies the cluster that a grid cell belongs to, $\phi$ $\epsilon$ $N$. The threshold $\tau$ is a fixed neighborhood radius, $\tau$ $\epsilon$ $N$. This parameter is similar to $\sigma(t)$ (neighborhood radius) used in the traditional version of SOM. The difference is that $\tau$ is modified to be a constant to ensure that the grid cells that are part of a cluster are adjacent to each other geographically.

\subsubsection{Determining the winning node} \label{section:SS-SOM_WinNode}
``Winning nodes'' are nodes assigned to a specific cluster in each learning cycle based on spatial and social attitude proximity. Therefore, a winning node, $n_{win}$ $\epsilon$ $S$, is defined as a node that: (a) is socially (based on prevalence of the social process) closest to the input vector $v$. In other words, among all the output nodes within the threshold $\tau$ of $n_{win}$, the difference between the weights of the winning node and the input vector $v$ is the least, and (b) for nodes that are equally close socially, the cluster $\phi$ that surrounds $n_{win}$ the most topologically (cluster $\phi$ with the maximum number of grid cells in radius $\tau$ of $n_{win}$), will influence $n_{win}$ most. 

To achieve this, at each iteration $t$ $\epsilon$ $(1,t_{max})$ we create a subset of grid cells, $\mu$, that are at a distance $\tau$ from $n_{win}$. Then, each grid cell $\mu_j$ that belongs to a cluster $\phi$ is denoted by $\mu_{j_{\phi}}$. Accordingly, the weight of $\mu_j$, denoted by $\omega_j$, is the same as the weight of all the grid cells in the same cluster, $\mu_{j_{\phi}}$, and the number of grid cells in cluster $\phi$ is $|\mu_{j_{\phi}}|$. Hence, as $\mu$ $\subseteq$ $S$, $\forall$ $\mu_j$ $\epsilon$ $\mu$,  $\mu_{win}$ is the same as $n_{win}$. Therefore, $\mu_{win}$ must satisfy the condition:
\begin{equation}
\label{eqn:04}
\forall \mu_j \epsilon \mu\text{, }||\omega_{win} - v||\text{ }|\mu_{j_{\phi}}| \leq  ||\omega_{j} - v||\text{ }|\mu_{j_{\phi}}|
\end{equation}
The difference is weighted by the size of the cluster in the radius $\tau$ ($|\mu_{j_{\phi}}|$) to maintain topological similarity. Then for SS-SOM, the best matching unit, BMU$(v)$, is calculated using the same formula as in traditional SOM, but by replacing the distance between a node $w_i$ and the input vector $v$ with the distance weighted by the size of the cluster in radius $\tau$.

\subsubsection{Weight Adaptation} \label{section:SS-SOM_weight}
In SS-SOMs, the weight adaptation step in traditional SOMs has been modified in accordance with the aim of giving equal importance to geographical as well as social distance. While the threshold $\tau$ and the BMU ensure we select adjacent grid cells,  $\omega_i(t+1)$ is adjusted to also take the proportion of a social process in cluster $\mu_{j_{\phi}}(t)$ into consideration. Additionally, the neighborhood function, $h(d,t)$ is modified \cite{kiviluoto1996topology,baccao2004geo,intelligence2012introduction} to ensure that the weight of each grid cell in the cluster, and the potential new grid cell are all considered:  $h(d,t) = e^{-\frac{d}{ |\mu_{j_{\phi}}(t)|+1}}$.

Finally, the weight of each output node is adapted as in the original SOM case, for $t$ $\epsilon$ $(1,t_{max})$:
\begin{equation}
\label{eqn:05}
\omega_{j}(t+1) = \omega_{j}(t) + (\eta(t)\text{ }h(||\mu_{win}-\mu_{j}||,t)\text{ }[v_j - \omega_{j}(t)]) 
\end{equation}


\section{Comparison of pipeline to baselines}
\subsection{Tweet Classification}
First, we validate the neural net text classification in the SS-SOM pipeline by comparing its performance to a baseline linear classifier (Support Vector Machines) which has been demonstrated as the most robust linear classifier for text classification \cite{JoachimsTextCategorization}.

\subsubsection{Support Vector Machines} \label{section:SVM}
Grid search was used to optimize hyper-parameters of the classifier selecting from a radial basis function kernel (gamma: 1e-3 to 1e-4) and linear kernels each with C values of \{1, 10, 100, 1000\} \cite{JoachimsTextCategorization}. Features (n-gram) used for the grid search ranged from one to six words, half the number of average words per Tweet \cite{greenTweitterAuthor}, coherent for comparison with the neural network approach.

\subsection{Administrative Boundaries and Clustering Methods}

We primarily compare the performance of resulting SS-SOM clusters with 2 baselines: (1) administratively defined Zip codes and (2) traditional version of SOM. We also discuss results of clustering by the Geo-SOM method \cite{baccao2004geo}.

\subsubsection{Zip codes}
\begin{figure}
  \includegraphics[trim=3cm 0cm 0cm 0cm, clip, width=\columnwidth]{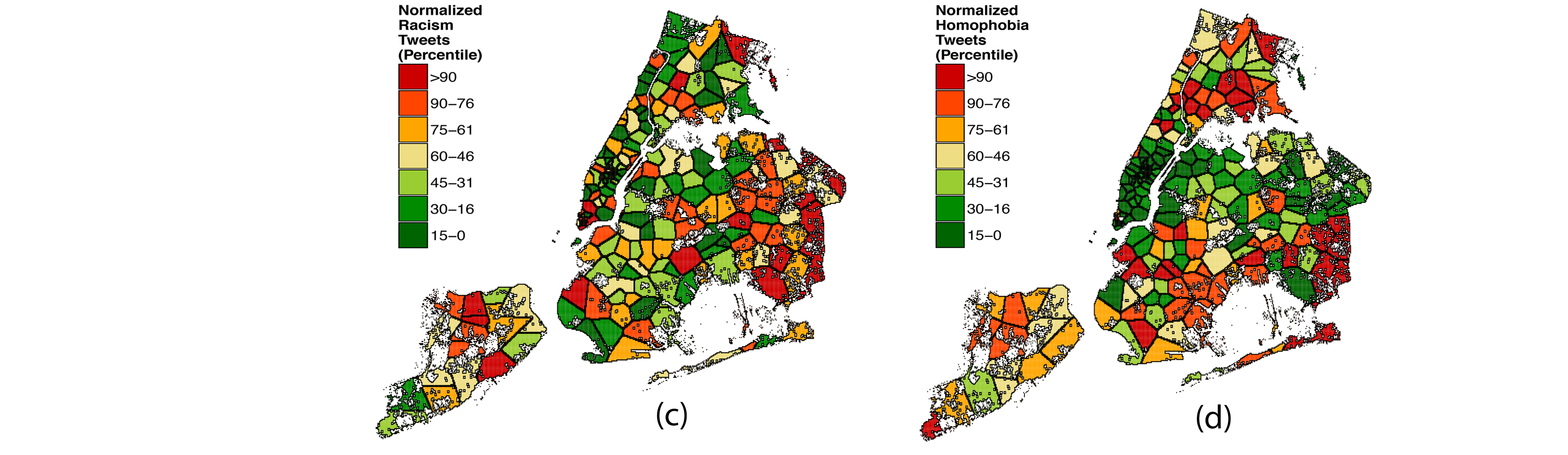}
  \caption{(a) Map of NYC depicting prevalence of social media measured racism by Zip code; (b) homophobia by Zip code; (c) racism by SS-SOM with 94 clusters (threshold of 3); (d) homophobia by SS-SOM with 102 clusters (threshold of 3). It is clearly visible that areas of high/low exposure are resolved differently using SS-SOMs versus Zip codes. The white-areas in (c) and (d) within the city represent grid cells that had 0 Tweets, hence normalization for such grid cells was not possible.}
  \label{fig:zipandSOMmap}
\end{figure}

Zip codes are generally used to quantify environmental exposures in health-related studies. To determine the prevalence of racism and homophobia in each Zip code, we mapped each grid cell to a Zip code using the NYC Zip code shapefile\footnote{https://www.census.gov/geo/maps-data/data/cbf/cbf\_zcta.html}. For each Zip code, we normalized the number of racism/homophobia Tweets with total number of Tweets in that Zip code. Figures \ref{fig:zipandSOMmap}(a) and \ref{fig:zipandSOMmap}(b) depict the prevalence of racism and homophobia Tweets by Zip codes.

\begin{figure}
  \includegraphics[trim=3cm 0cm 0cm 0cm, clip, width=\columnwidth]{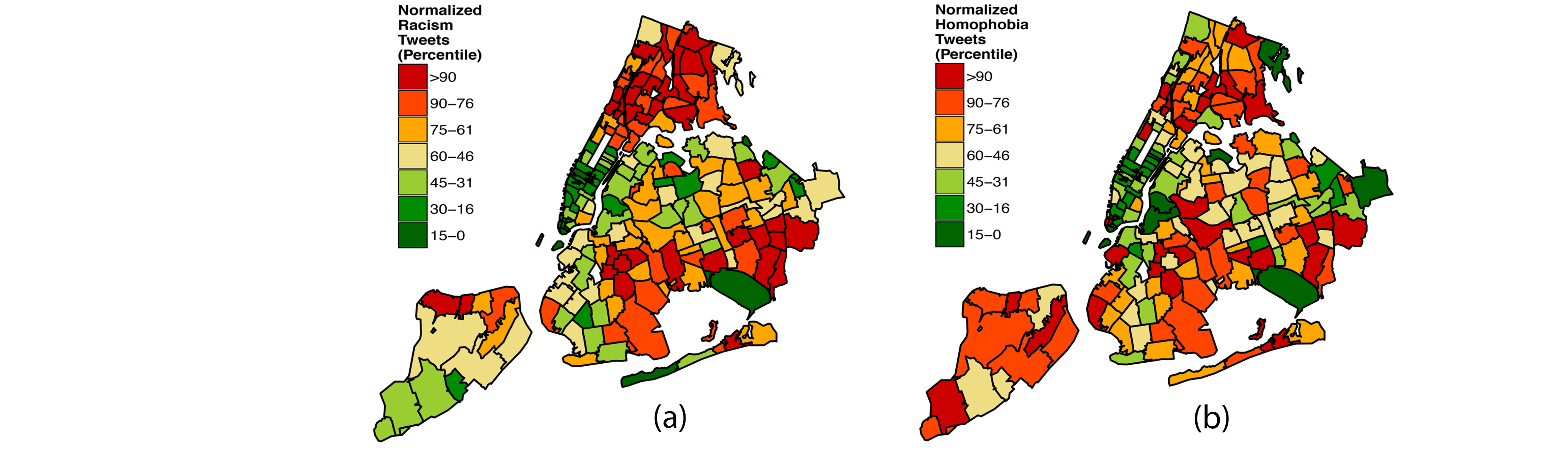}
  \end{figure}

\subsubsection{Traditional Self-Organizing Maps} \label{section:tradSOM}
We define the following input variables (same as in SS-SOM): $g$, the number of grid cells; and $\mathbf{I}$, an input vector space having four vectors: number of Tweets in a grid cell, number of racism/homophobia Tweets, and the latitude and the longitude of the grid \cite{baccao2004geo}; length of $\mathbf{I}$ = $g$. The operational parameters used are in line with previous work \cite{intelligence2012introduction}. Implementation led to 3622 clusters for racism, and 2899 clusters for homophobia. We used a maximum $t_{max}$ of 17500 as used in \cite{giraudel2001comparison} thus decreasing the number of clusters by only 20\% to 2899 for racism and 2328 for homophobia which was still well above the number from SS-SOM which represent a number of geographic areas of exposure that are more comparable to Zip codes.

\subsubsection{Geo-SOM} \label{section:GeoSOM}
The other closest baseline algorithm we can compare to is the Geo-SOM algorithm \cite{baccao2004geo}. The aim of this work, to develop homogeneous regions and perform geographic pattern detection by taking into account spatial dependency along with social dependency, was achieved by changing the neighborhood function to search for BMU only within a predefined topological vicinity. While such an approach guarantees homogeneous clusters, we implemented this algorithm using the software (Geo-SOM Suite)\footnote{http://www.novaims.unl.pt/labnt/geosom/GeoSOM.suite.htm} to confirm that it does not fulfill our constraints regarding generation of contiguous and non-overlapping partitioning of the city and hence, we do not further evaluate this baseline with SS-SOM.

\section{Evaluation}
First, we evaluate the text classification performance (for identifying racism or homophobia Tweets). Then, we evaluate clustering quality of the SS-SOMs. Finally, we evaluate how meaningful resulting SS-SOMs are by examining how online homophobia/racism exposure measured through SS-SOMs differs from what would be measured through Zip codes, and qualitatively examining the resulting prevalence of racism and homophobia.

\subsection{Classification Performance}
To evaluate performance of racism and homophobia classification, we compare the results of neural network classification with our baseline (SVM) based on the F1 score and Receiver operating characteristic (ROC) curve from each method.

\subsection{Cluster Quality} We evaluate implementations that result in clusters that geographically segment NYC (SS-SOM, Zip codes and SOM)  along the following 4 axes related to the cluster quality: (1) \textbf{Similarity} indices to quantify how grid cells land in clusters across methods, thus conveying how different the overall clustering in NYC is by SS-SOM and the baseline methods. (2) \textbf{Robustness of results to missing data} for each method. Specific to the application here, missing data may arise from an under-represented area on social media or inconsistent data from Twitter API. Accordingly, we evaluate sensitivity to missing data in two different ways: (i) missing Tweets and (ii) missing grids. (3) \textbf{Mean variance} of social processes across grids in resulting clusters. This directly assesses homogeneity of the resulting clusters; how internally similar the resulting clusters are. (4) \textbf{Mean Squared Prediction Error (MSPE)} is used to assesses reliability of the clusters by calculating the mean squared difference of the prevalence of racism/homophobia between clusters formed using complete versus held-out data. 


\subsubsection{Similarity of SS-SOM to Baselines} As using the SS-SOM method results in a different number of clusters compared to Zip codes and traditional SOM method (and also depends on the threshold selected for the SOMs), we compare the clustering between methods using the similarity index $c2(Y,Y')$ \cite{Rand1971clusteringEval}. This is the ratio of the number of similar assignments of (grid) points to the total number of grid point pairs. The higher the value of $c2$, the more similar the resulting clusters from the comparing methods are. More precisely, given $n$ points: $x_1, x_2,..., x_n$, and clustering results: $Y = {Y_1,...,Y_{k1}}$ and $Y' = {Y'_1, ... , Y'_{k2}}$, $c2(Y,Y') = \frac{\sum\limits_{i<j}^{N}\gamma_{ij}}{\binom{N}{2}}$ where $\gamma_{ij} =$ 1 if there exist $k$ and $k'$ such that both $x_i$ and $x_j$ are in both $y_k$ and $y'_{k'}$ or if there exist $k$ and $k'$ such that both $x_i$ is in both $y_k$ and $y'_{k'}$ and $x_j$ is in neither $y_k$ or $y'_{k'}$, or 0 otherwise.

\subsubsection{Robustness to Missing Data} For missing Tweets, we used a holdout range of $n_1$=\{25\%, 50\% and 75\%\} of the total number of Tweets and assessed the resulting clusters formed by traditional SOM and SS-SOM. For missing grids, we used an approach from clustering evaluation \cite{Rand1971clusteringEval}. We compare traditional SOM and SS-SOM regarding missing grid sensitivity, as Zip code boundaries don't change if data is missing. We first perform the clustering using a subset of $n_1$ grids. Then, we implement the same clustering methods on additional $n_2$ $= n - n_1$ grids (for a total $n = n_1 + n_2$ grids). We compare how the original $n_1$ grids are clustered using the larger data set ($n$) using a k-fold cross validation ($k$=10) for $n_1$ (and $n$ = 72,484 grid cells) and similarity index $c2$. 
\subsubsection{Variance of Social Processes by Cluster} Mean variance of grid proportions in each cluster is: $s^2 = \frac{1}{n-1} \sum\limits_{i=1}^n (x_i - \bar{x})^2$ where $n$ is the number of grid cells in a cluster, $x_i$ is the normalized homophobia/racism count of the $i$th grid cell and $\bar{x}$ the normalized homophobia/racism Tweet count in the entire cluster. 

\subsubsection{Mean Squared Prediction Error (MSPE)} MSPE for SS-SOM is calculated by holding out 10\%, 25\%, 50\% and 75\% of data (we performed this for both missing grid cells, and missing Tweets separately). We re-learn the spatial model for each size of the data, disregarding the ``weight'' for the held out grid cells, and compute the prevalence of racism/homophobia for the newly formed clusters. We then compute the MSPE using the prevalence of racism/homophobia in the clusters each of those grid cells is assigned to in each case above. We do a $k$-fold cross-validation ($k$=10) to prevent over-fitting. $MSPE$ is computed by: $E\big[\sum\limits_{i=1}^{\text{n}}(g(x_i) - \hat{g}(x_i))^2\big]$
where, $n$ are the number of grid cells in the held out data, $g(x_i)$ is the mean prevalence of the cluster grid cell $i$ is assigned to in the full model, and $\hat{g}(x_i)$ when it is held out.

\subsection{Cohort's Exposure to Racism and Homophobia via SS-SOM vs. Zip codes}
We calculated exposure differences ($E_{i}$), for each person $i$ in the P18 cohort by assessing the difference in proportion of racism/homophobia Tweets for each cell visited by individual $i$, as a proportion of all their visits. 
\begin{equation}
\label{eqn:06}
E_i = \bigg(\sum\limits_{k=1}^{\text{n}} \big(\frac{\left|E_{zip_{k}}-E_{SOM_{k}}\right|}{E_{zip_{k}}} * V_{i_{k}}\big)\bigg)/V_i
\end{equation}

where, $E_{zip_{k}}$ is the proportion of racism or homophobia Tweets in Zip code containing grid cell $k$ and $E_{SOM_{k}}$ is the proportion of racism or homophobia Tweets in SS-SOM cluster containing grid cell $k$. $V_{i_{k}}$ is the number of visits made to grid cell $k$ by person $i$ and $V_i$ is the total number of visits made in entire NYC by person $i$.

\subsection{Prevalence and Co-prevalence of Racism and Homophobia Measured Through Social Media}
We know methodologically that Zip codes will suffer from spatially averaging of the racism and homophobia sentiments, but will this result in differing prevalence of racism and homophobia online sentiment in NYC? Here we demonstrate the impact of results by qualitatively examining the prevalence of racism and homophobia sentiment through the Zip code and SS-SOM measures.


\section{Results}

\begin{figure}
\includegraphics[trim=2.5cm 5cm 0cm 2.5cm, clip, scale=0.5]{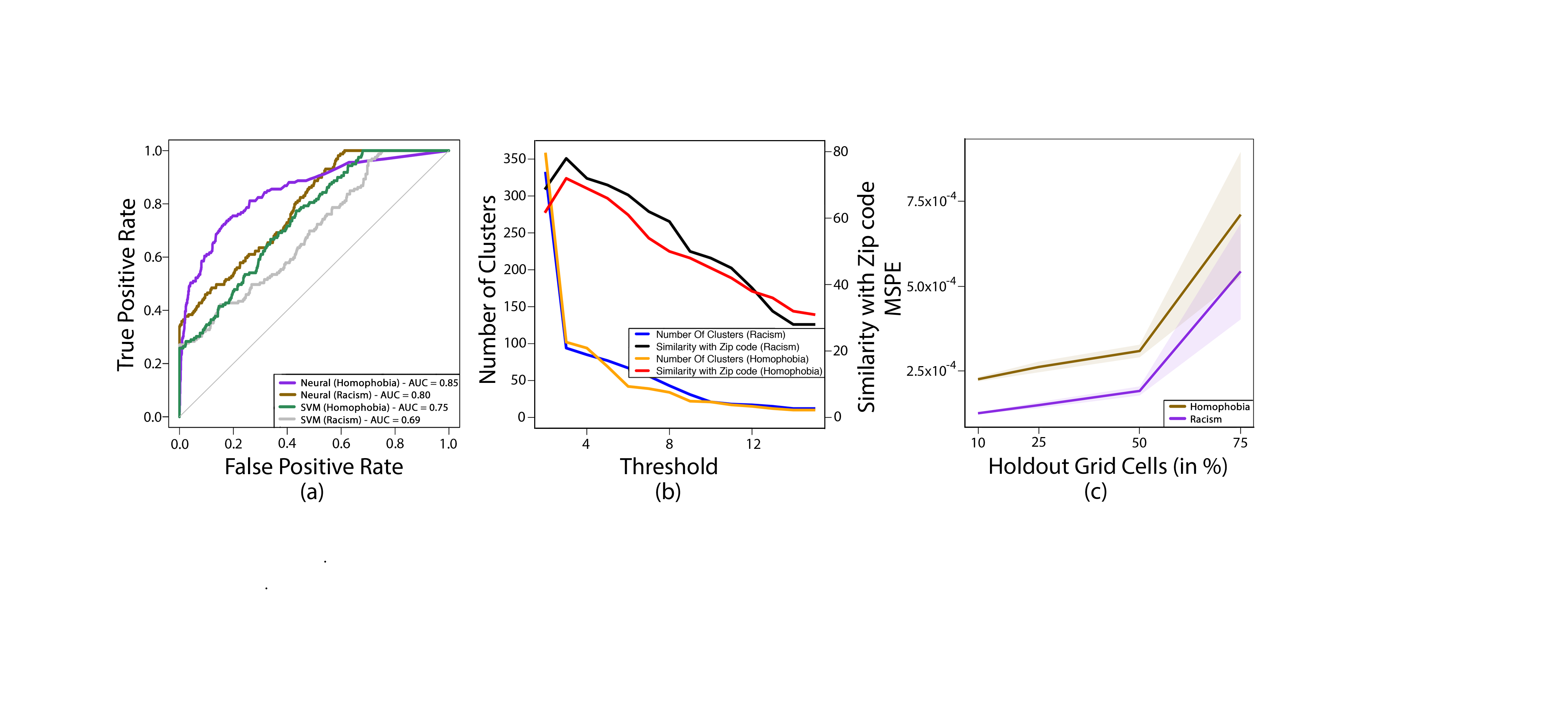}
  \caption{(a) ROC curve for racism and homophobia classification using neural embedding and SVM. (b) Number of clusters and similarity between Zip codes and SS-SOM by threshold value. (c) Mean squared prediction error (MSPE) for SS-SOM (threshold of 3) across different proportions of held out grid cells (shaded region represents 95\% confidence interval).}
  \label{fig:similarity}
\end{figure}


\subsection{Classification Performance}
Neural embedding for racism classification (F1 score: 0.89, Area Under Curve (AUC): 80) (Figure \ref{fig:similarity}a) was slightly better than SVM (F1: 0.85, AUC: 0.69). This was also the case for homophobia classification; F1 scores of 0.89 and 0.86 respectively and AUC of 0.85 and 0.75 respectively \cite{tang2014learning}. We found that the neural embedding classifier yields a low false positive rate (precision for racism-0.89, homophobia-0.87) and low false negative rate (recall for racism-0.80, homophobia-0.84) on test data. Therefore, we can conclude that it is efficiently able to classify Tweets into racism versus no racism and homophobia versus no homophobia.

\subsection{Cluster Quality}

\subsubsection{Similarity of SS-SOM to Baselines} \label{section:Result_similarity}
For $n$=72,484 grid cells, we find that Zip code and SS-SOM methods are most similar in terms of homophobia and racism by cluster (72\% and 78\% respectively). A comparison between homophobia and racism clusters can only be assessed via similarity metrics for the SS-SOM method (82\%) showing a high similarity between clustering of the two processes. The traditional SOM is least similar to both Zip codes (25\% and 28\% for homophobia and racism respectively) and SS-SOM (34\% and 36\% for homophobia and racism respectively), due to the high number of clusters obtained (3622 clusters for racism, and 2899 clusters for homophobia). 

Figure \ref{fig:similarity}b shows $c2$ versus threshold value. Similarity between SS-SOM and Zip codes for both homophobia and racism peaks at a threshold of 3 which also has a reasonable balance between amount of Tweets per cluster and computation time (as discussed in section \ref{section:SS-SOM_desc}). Thus all further analyses related to SS-SOM use a threshold of three.

\subsubsection{Sensitivity to Missing Data}
\label{section:missing_result}
Missing Tweets resulted in no effect in both SOM and SS-SOM. This is because for each grid, we used a normalized proportion of Tweets. As racism/homophobia Tweets are just as likely to be missing as any other Tweets, normalization accounts for missing data. For missing grid cells, a similarity ($c2$) of up to 95.3\% and 99.3\% (SOM and SS-SOM respectively) between the clustering from the original $n$ and $n_1$ grid cells. Overall, for a 10-fold cross validation for 25\%, 50\% and 75\% of missing grid cells, the average similarity for traditional SOM was 95.3\%, 89.2\% and 82.8\% respectively (standard deviation up to 1.2\%), and the average similarity for SS-SOM was 99.3\%, 93.8\% and 85.8\% (standard deviation of up to 1.3\%) which indicate that SS-SOMs are generally stable with respect to missing data.

\subsubsection{Variance of Social Processes by Cluster} \label{sec:VarianceResult} Traditional SOM and SS-SOM had lower mean variance for racism clusters than Zip codes (Table \ref{tab:meanvar}), and all methods were similar for homophobia clusters. Notably, minimums for both racism and homophobia by Zip codes are zero for Zip code 10470 (northern Bronx). This is because only 85 tweets from the entire data set were located in this Zip code, and none were racist or homophobic. Both SS-SOM and traditional SOM methods create a more consistent set of clusters than Zip codes (in terms of variance), thus improving on Zip codes for creating social-process segmented regions. Though the traditional SOM had the least variance leading to more homogeneous clusters, there are two reasons which make the use of traditional SOM not feasible: (a) the high number of clusters that the city gets divided into, and (b) the clusters are not topologically constrained and also overlap, which are basic requirements for the problem.

\begin{table}%
\caption{Mean variance of social process by clustering technique. Traditional SOM shows the lowest variance for racism and homophobic-specific clusters.}
\label{tab:meanvar}
\begin{minipage}{\columnwidth}
\begin{center}
\begin{tabular}{l |l|l}
  \toprule
    {\small \textbf{Clustering}}
    & {\small \textbf{\begin{tabular}[c]{@{}l@{}}Racism Mean Variance (Min, Max)\end{tabular}}}
    & {\small \textbf{\begin{tabular}[c]{@{}l@{}}Homophobia Mean Variance (Min, Max) \end{tabular}}}
       \\
       \hline
    \textbf{Zip code} & 1.1x$10^{-3}$ (0,0.029) &  1.7x$10^{-3}$ (0,0.013)\\
    \hline
    \textbf{SOM} & 9.1x$10^{-7}$(2.3\textnormal{x}$10^{-8}$,7.2\textnormal{x}$10^{-4}$) &  6.2x$10^{-7}$ (8.3\textnormal{x}$10^{-9}$,1.4\textnormal{x}$10^{-5}$)\\
    \hline
    \textbf{SS-SOM}&9.4x$10^{-4}$ (1.6\textnormal{x}$10^{-6}$,6.3\textnormal{x}$10^{-3}$) &  2.9\textnormal{x}$10^{-3}$ (4.4\textnormal{x}$10^{-6}$,1.4\textnormal{x}$10^{-3}$) \\
  \bottomrule
\end{tabular}
\end{center}
\bigskip\centering
\end{minipage}
\end{table}%

\subsubsection{Mean Squared Prediction Error (MSPE)}

For grid cells, MSPE values increased from 1.25x10$^{-4}$ to 5.4x10$^{-4}$ (racism) and 2.3x10$^{-4}$ to 7.1x10$^{-4}$ (homophobia) from 10\% to 75\% of grid cells held out (Figure \ref{fig:similarity}(c)). The low MSPE values suggest that the clusters created by SS-SOM are homogeneous. The MSPE for SS-SOM after holding out Tweets instead of grid cells is nearly zero (maximum MSPE is 7.7x10$^{-5}$ for 75\% holdout Tweets) across all hold-out sizes; this is  because SS-SOMs are robust to missing Tweets due to normalization. Finally, MSPE of traditional SOM and Zip codes are not calculated as the former doesn't meet our clustering requirements as discussed at the end of section \ref{sec:VarianceResult} and the boundaries of the latter don't change.

\subsection{Cohort's Exposure to Online Racism and Homophobia}
There was a clear difference in the cohort's measured exposure to online racism and homophobia for SS-SOM versus Zip codes. Mean racism exposure difference was 41.3\% (SD: 17.8\%). For 35\% (78/226) of individuals, the difference was over 50\%. Mean homophobia exposure difference was 29.6\% (SD: 18.7\%). Qualitatively we can identify places with large differences in exposure that would be important based on how many times they are visited; a grid cell which had 71,903 visits in upper Manhattan (near the Bronx border) had the highest difference in racism Tweet prevalence through Zip codes (2.2\%) and SS-SOM (0.7\%). A grid cell (which had 109,876 visits) near the City Hall in Lower Manhattan had the highest difference in homophobia Tweet prevalence through Zip codes (3.0\%) and SS-SOM (1.0\%).

\subsection{Prevalence and Co-prevalence of Racism and Homophobia Measured Through Social Media}
The maps in Figure \ref{fig:zipandSOMmap} show prevalence of racism and homophobia in Zip codes and SS-SOM clusters. For Zip codes, the proportion of racism Tweets in each Zip code ranged from 0\% to 3.60\% and the proportion of homophobia Tweets ranged from 0\% to 3.20\%. For the SS-SOM clusters, the proportion of racism Tweets in each cluster ranged from 0.28\% to 1.23\% and the proportion of homophobia Tweets ranged from 0.09\% to 2.92\%. An important observation about the ranges for proportion by Zip code is that extreme values result for both racism and homophobia due to a very few number of Tweets being made in some Zip codes. For example, the Zip code with the lowest prevalence of racism just had 306 Tweets, and the Zip code with the lowest prevalence of homophobia just had 71 Tweets. The SS-SOM cluster with the lowest prevalence of racism had 100,774 Tweets and the SS-SOM cluster with the lowest prevalence of homophobia had 16,365 Tweets.

\begin{figure}
  \includegraphics[trim=4cm 2.5cm 0.5cm 3.5cm, clip, scale=0.5]{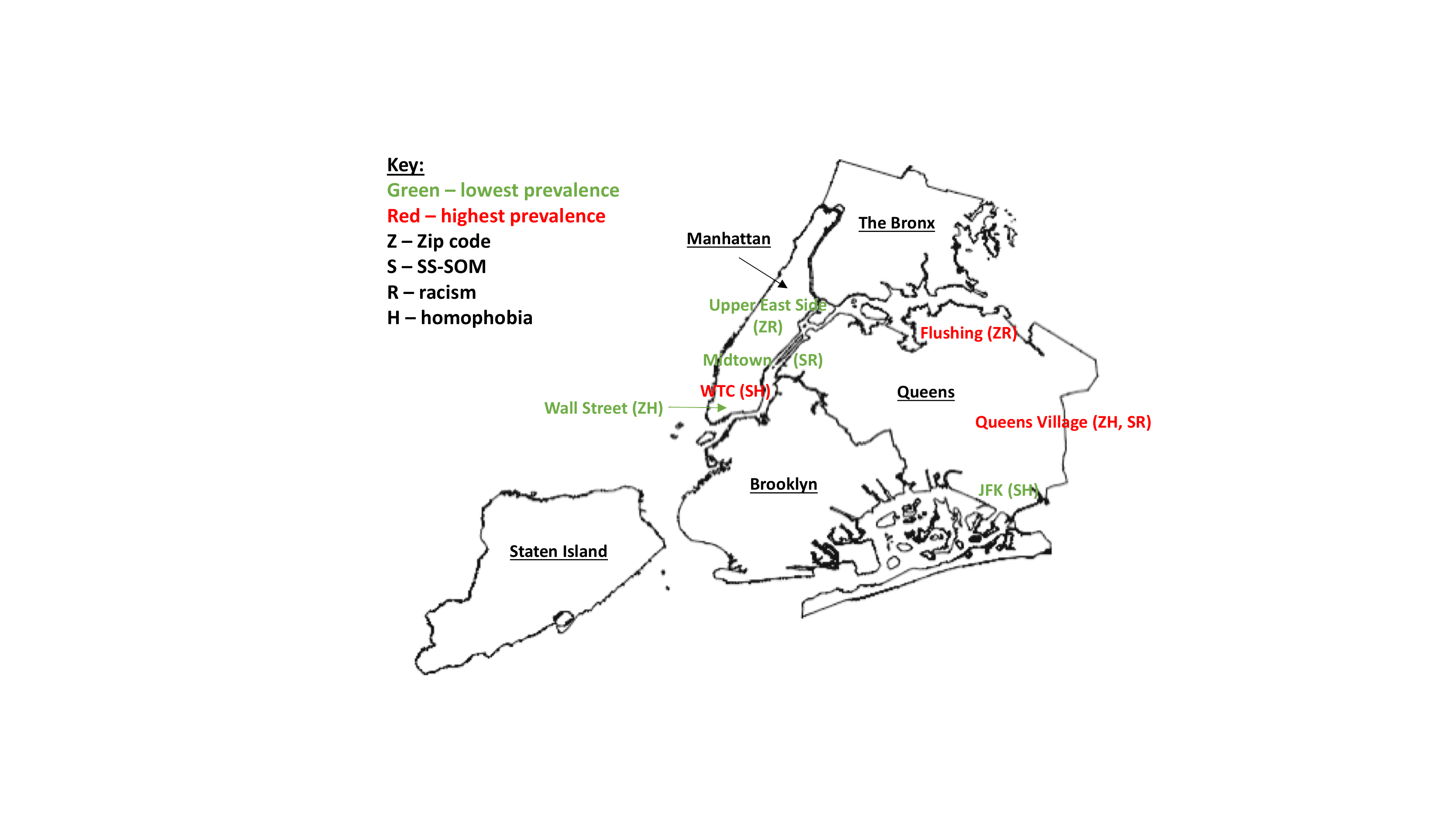}
  \caption{NYC map displaying areas of online homophobia/racism prevalence for geographic reference.}
  \label{fig:NYCMap}
\end{figure}

Qualitatively, we see that differences from these methods result in some overall similarities and differences. For racism, both Zip codes and SS-SOMs show high prevalence in areas of Staten Island and Queens. However, one striking difference is in the Bronx where most of the Zip codes show high prevalence, but many of the SS-SOMs show low prevalence. We examined the proportion of racist Tweets in this area and found that many Zip codes appear to have a high proportion of racism when it is mostly concentrated in one or two grid cells which can be adjacent to grid cells with much lower prevalence. These grids get split into different SS-SOMs, which internally will be more consistent, and overall the proportion in all of those SS-SOMs will be lower. Thus, the proportion of racism in Zip codes is being mis-represented due to a high proportion in one or two grid cells. Similarly for homophobia, the neighborhood with highest homophobia online  via Zip codes is in Queens Village, however for SS-SOM it is near the World Trade Center (Figure \ref{fig:NYCMap}), due to a similar spatial averaging effect. Both methods show a relatively low level of racism and homophobia in Manhattan, though SS-SOMs define some areas with higher racism in Manhattan.


Comparing the SS-SOM racism and homophobia clusters, we found a similarity of 82\% (as discussed in \ref{section:Result_similarity}). Some qualitative differences were in northern Staten Island (Mariners Harbor) (higher racism but below average homophobia) and near the world trade center (WTC) which had higher homphobia but below average racism. Overall we found that there were a lot of clusters that ranked low in racism but were very high in homophobia (like Rockaway Park Queens, Lower Manhattan near the World Trade Center). This indicates that the prevalence of online racism is generally accompanied by prevalence of homophobia but the opposite is not true; areas may have higher levels of homophobia independent of racism. This also holds true partially due to a higher prevalence of homophobia on Twitter in NYC as compared to racism, hence prevalence of homophobia may not always be accompanied by racism. We also analyzed the co-prevalence of racism and homophobia at a user level and found the same result: only 2.5\% of the users with a homophobia Tweet also had a racism one, but the reverse was 10\%. 


\section{Discussion and Conclusion}
\subsection{Summary of Findings and Contributions}
SS-SOMs are a simple modification of SOMs that address the need for partitioning a complete area into non-overlapping, contiguous and homogeneous areas that best represent the latent sentiment behind sparse and noisy Tweets, while accounting for varying social media levels in different places. This work adds to the social computing literature using novel data sources to augment our understanding of spatial regions \cite{intagorn2011learning,Rattenbury2009placeFlickr,hollenstein2010exploring}. Overall, we found that SS-SOMs show robustness to missing data, lower maximum variance in cluster content and less evidence of spatial averaging as compared to Zip codes. Mean variance of homophobia/racism in SS-SOM clusters was similar to that of traditional SOM (lower maximum variance) and lower than variance of Zip codes. Both SOM and SS-SOM had a narrower range of variances across all clusters compared to Zip codes, indicating more consistency. Thus, we see that the spatial averaging resulting by using Zip code measures of online racism and homophobia would be significant compared to SS-SOMs, for a real population for which racism and homophobia are relevant. We further found that SS-SOM had low MSPEs across a range of proportions of grid cells missing, indicating good reliability.

This work creates a novel approach to understand underlying sentiment in areas based on individual, sparse and noisy social media data. This is useful for topics that are difficult to measure in other ways, like racism or homophobia. It has been shown that people are vocal about such sensitive subjects on social media where they perhaps feel more comfortable than on surveys or other face-to-face situations \cite{williams2015cyberhate}. Therefore, there are no ``gold standard'' methods that capture the same types of prejudice as found on social media (e.g. hate crime reports capture prejudice that results in crimes), and can be exactly compared and used to validate the findings here. Instead, social media measures of prejudice should be seen as a complement to other measures. The thorough qualitative examination of the identified  areas of racism and homophobia, as well as assessment of resulting exposure to a real cohort, confirms the importance of developing these measures and ways to operationalize the use of the social media data and compare and integrate it with other measures of prejudice.

In comparison of SS-SOM and traditional SOM to Zip code, SS-SOM clusters were more similar to Zip code for both racism and homophobia compared to similarity between traditional SOM and Zip codes. This is not surprising given that the traditional SOM clusters are high in number as compared to Zip code. However, both methods had lower similarity with Zip codes for racism (compared to homophobia) indicating racism may have determinants more unrelated to Zip code-defined areas. We also found that an increase in racism was generally associated with homophobia, but not the opposite. Thus this work can also be used to develop studies to assess the reason behind levels or co-prevalence of these social processes. 

Finally, we demonstrated via mobility of a cohort of MSM (for whom exposures to homophobia and racism are important health risk factors), that exposures measured via Zip codes and SS-SOM can be largely different (over 50\% different for racism exposure in a third of the cohort). This is an important finding; exposure measures linked to Zip codes suffer from averaging whereas SS-SOM clusters may represent the underlying exposure more consistently; SS-SOM clusters that were high or low for one social process were in regions that had average or opposite levels based on Zip code or traditional SOM definitions.

\subsection{Limitations and Future Work}
An inherent limitation in this work is the use of the Twitter API. Though results are robust to missing data so amount of data should not be an issue, any inherent biases in the API sample (e.g. by content, location, language) would affect the results. As well, augmented text processing methods such as including sentiment could also be explored in order to more optimally encompass the diversity of human language; we used a fairly simple approach to text classification via n-grams.

An important next step should assess how the sentiment measured through the online environmental measures connects to the offline world (it should be noted this is not the same as linking a Tweet to the \emph{home} location of a user -- doesn't exactly correspond to the ``localness assumption'' \cite{johnson2016geography}). As discussed, the literature indicates there is a need for better and new racism and homophobia measures, which motivates this work. In some way, relationships between online data and outcomes have been demonstrated by work examining links between area-based measures of racial animus from Google searches and black mortality outcomes \cite{Chae2015blackMortality} or differential votes for a black candidate \cite{stephens2014cost}. As well, this assumption is used in many studies which harness geo-location of social media and offline attributes such as in assessing place semantics or sound maps \cite{Rattenbury2009placeFlickr,aiello2016chatty}. However a systematic study for our measures and health outcomes, such as through using surveys (which are currently not available for such social processes) could be performed. As described above, racism/homophobia measures from social media may be different from self-reported racism sentiment or experience on surveys, so this study would have to be carefully designed, perhaps using focus groups with select social media users. Further analytic work can be done to tease apart the types of racism/homophobia online (e.g. expressions versus reports of encounters), which may have different spatial patterns. As well, since some people are more vocal in expressing than others, a study should examine if ``louder'' users on Twitter correspond to an increase in offline sentiment. 

As social media has high temporal resolution and is available in real-time, future efforts could also explore how clusters could be monitored over time to assess changes in exposures. Although we specifically selected NYC for this study based on the P18 data, and the understanding that racism and homophobia can be important social determinants for groups like MSM in urban areas like NYC, the generalizability and utility of this work in other cities and non-urban areas should be studied. Finally, our finding that the prevalence of online racism is generally accompanied by prevalence of homophobia but the opposite is not true should be investigated for its sociological underpinnings.

\subsection{Implications and Design of Social Computing Systems}
The SS-SOM pipeline can be of use for researchers and practitioners when exploring the geographical properties of social processes, especially those that may uniquely be communicated over social media. In order to facilitate appropriate interventions, many studies have been done to understand the prevalence of hate speech in the online world \cite{Chaudhry2015,silva2016analyzing,bartlett2014anti,burnap2015cyber}. From this work, areas defined by specific social topics can be used to assess relevant social influence or context for an individual or to control (normalize) for their social environment. System design, actively deploying interventions, or generation of feedback based on an individual's social environment can also be better focused based on a contextual measure of neighborhood. For example, approaches to mitigating divisive sentiment at the right place and time can be targeted based on knowledge of spatial patterns of these social processes.


\appendix
\section{List of Keywords}
\textbf{Racism keywords:} nigger, niggers, racism, racist, coon, ape, abid, abeed, ali baba, alligator bait, beaner, beaney, bootlip, buffie, burrhead, ching chong, chinaman, abcd, coolie, groid, haji, hajji, sooty, spic, spook, tar-baby, toad, wigger, ``white nigger'', white trash, ``white trash''.

\hspace{-0.35cm}\textbf{Homophobia keywords:}``batty boy'', dyke, faggot, ``fag bomb'', ``gay mafia'', ``gold star gay'', ``gold star lesbian'', ``no homo'', queer, twink, ``kitty puncher'', ``brownie king'', twink, gaysian, ``gym bunny'', daffy.

\section{AMT Instructions} The racism labelling task instructions are given as an example, homophobia instructions were the same with appropriate racism/homophobia changes: ``We are interested in labeling Tweets that exhibit racism or exposure to it. So, please label a Tweet as positive only if: 1) it exhibits a prejudice against any race (e.g. black, hispanic, asian, etc.), 2) it has a negative/derogatory remark about any race (e.g. black, hispanic, asian, etc.), and 3) it indicates that the person Tweeting was exposed to racism or observed someone being exposed to racism.''
These instructions were followed by examples of racism and non-racism Tweets and explicit discussion of special cases (e.g. song lyrics or colloquial terms); through iteration we found specific direction about these was needed.

\begin{acks}
The authors acknowledge support by the National Institutes of Health and National Institute of Mental Health (NIMH) under grant R21 MH110190, and by the National Science Foundation under award MRI-1229185.
\end{acks}


\bibliographystyle{ACM-Reference-Format}
\bibliography{CSCW_bibliography}


\begin{thebibliography}{69}


\ifx \showCODEN    \undefined \def \showCODEN     #1{\unskip}     \fi
\ifx \showDOI      \undefined \def \showDOI       #1{#1}\fi
\ifx \showISBNx    \undefined \def \showISBNx     #1{\unskip}     \fi
\ifx \showISBNxiii \undefined \def \showISBNxiii  #1{\unskip}     \fi
\ifx \showISSN     \undefined \def \showISSN      #1{\unskip}     \fi
\ifx \showLCCN     \undefined \def \showLCCN      #1{\unskip}     \fi
\ifx \shownote     \undefined \def \shownote      #1{#1}          \fi
\ifx \showarticletitle \undefined \def \showarticletitle #1{#1}   \fi
\ifx \showURL      \undefined \def \showURL       {\relax}        \fi
\providecommand\bibfield[2]{#2}
\providecommand\bibinfo[2]{#2}
\providecommand\natexlab[1]{#1}
\providecommand\showeprint[2][]{arXiv:#2}

\bibitem[\protect\citeauthoryear{Aiello, Schifanella, Quercia, and
  Aletta}{Aiello et~al\mbox{.}}{2016}]%
        {aiello2016chatty}
\bibfield{author}{\bibinfo{person}{Luca~Maria Aiello}, \bibinfo{person}{Rossano
  Schifanella}, \bibinfo{person}{Daniele Quercia}, {and}
  \bibinfo{person}{Francesco Aletta}.} \bibinfo{year}{2016}\natexlab{}.
\newblock \showarticletitle{Chatty maps: constructing sound maps of urban areas
  from social media data}.
\newblock \bibinfo{journal}{\emph{Open Science}} \bibinfo{volume}{3},
  \bibinfo{number}{3} (\bibinfo{year}{2016}), \bibinfo{pages}{150690}.
\newblock


\bibitem[\protect\citeauthoryear{Aosved, Long, and Voller}{Aosved
  et~al\mbox{.}}{2009}]%
        {Aosved2009intolerence}
\bibfield{author}{\bibinfo{person}{Allison~C Aosved},
  \bibinfo{person}{Patricia~J Long}, {and} \bibinfo{person}{Emily~K Voller}.}
  \bibinfo{year}{2009}\natexlab{}.
\newblock \showarticletitle{Measuring sexism, racism, sexual prejudice, ageism,
  classism, and religious intolerance: The intolerant schema measure}.
\newblock \bibinfo{journal}{\emph{Journal of Applied Social Psychology}}
  \bibinfo{volume}{39}, \bibinfo{number}{10} (\bibinfo{year}{2009}),
  \bibinfo{pages}{2321--2354}.
\newblock
\showISSN{1559-1816}


\bibitem[\protect\citeauthoryear{Ba{\c{c}}{\~a}o, Lobo, and
  Painho}{Ba{\c{c}}{\~a}o et~al\mbox{.}}{2004}]%
        {baccao2004geo}
\bibfield{author}{\bibinfo{person}{Fernando Ba{\c{c}}{\~a}o},
  \bibinfo{person}{Victor Lobo}, {and} \bibinfo{person}{Marco Painho}.}
  \bibinfo{year}{2004}\natexlab{}.
\newblock \showarticletitle{Geo-self-organizing map (Geo-SOM) for building and
  exploring homogeneous regions}.
\newblock \bibinfo{journal}{\emph{Geographic Information Science}}
  (\bibinfo{year}{2004}), \bibinfo{pages}{22--37}.
\newblock


\bibitem[\protect\citeauthoryear{Bartlett, Reffin, Rumball, and
  Williamson}{Bartlett et~al\mbox{.}}{2014}]%
        {bartlett2014anti}
\bibfield{author}{\bibinfo{person}{Jamie Bartlett}, \bibinfo{person}{Jeremy
  Reffin}, \bibinfo{person}{Noelle Rumball}, {and} \bibinfo{person}{Sarah
  Williamson}.} \bibinfo{year}{2014}\natexlab{}.
\newblock \showarticletitle{Anti-social media}.
\newblock \bibinfo{journal}{\emph{Demos}} (\bibinfo{year}{2014}),
  \bibinfo{pages}{1--51}.
\newblock


\bibitem[\protect\citeauthoryear{Bor, Venkataramani, Williams, and Tsai}{Bor
  et~al\mbox{.}}{2018}]%
        {bor2018police}
\bibfield{author}{\bibinfo{person}{Jacob Bor}, \bibinfo{person}{Atheendar~S
  Venkataramani}, \bibinfo{person}{David~R Williams}, {and}
  \bibinfo{person}{Alexander~C Tsai}.} \bibinfo{year}{2018}\natexlab{}.
\newblock \showarticletitle{Police killings and their spillover effects on the
  mental health of black Americans: a population-based, quasi-experimental
  study}.
\newblock \bibinfo{journal}{\emph{The Lancet}} (\bibinfo{year}{2018}).
\newblock


\bibitem[\protect\citeauthoryear{Brauer}{Brauer}{2012}]%
        {intelligence2012introduction}
\bibfield{author}{\bibinfo{person}{Christoph Brauer}.}
  \bibinfo{year}{2012}\natexlab{}.
\newblock \showarticletitle{An Introduction to Self-Organizing Maps}.
\newblock  (\bibinfo{year}{2012}).
\newblock


\bibitem[\protect\citeauthoryear{Burnap and Williams}{Burnap and
  Williams}{2015}]%
        {burnap2015cyber}
\bibfield{author}{\bibinfo{person}{Pete Burnap} {and}
  \bibinfo{person}{Matthew~L Williams}.} \bibinfo{year}{2015}\natexlab{}.
\newblock \showarticletitle{Cyber hate speech on twitter: An application of
  machine classification and statistical modeling for policy and decision
  making}.
\newblock \bibinfo{journal}{\emph{Policy \& Internet}} \bibinfo{volume}{7},
  \bibinfo{number}{2} (\bibinfo{year}{2015}), \bibinfo{pages}{223--242}.
\newblock


\bibitem[\protect\citeauthoryear{Carpiano, Kelly, Easterbrook, and
  Parsons}{Carpiano et~al\mbox{.}}{2011}]%
        {carpiano2011community}
\bibfield{author}{\bibinfo{person}{Richard~M Carpiano},
  \bibinfo{person}{Brian~C Kelly}, \bibinfo{person}{Adam Easterbrook}, {and}
  \bibinfo{person}{Jeffrey~T Parsons}.} \bibinfo{year}{2011}\natexlab{}.
\newblock \showarticletitle{Community and drug use among gay men: The role of
  neighborhoods and networks}.
\newblock \bibinfo{journal}{\emph{Journal of Health and Social Behavior}}
  \bibinfo{volume}{52}, \bibinfo{number}{1} (\bibinfo{year}{2011}),
  \bibinfo{pages}{74--90}.
\newblock


\bibitem[\protect\citeauthoryear{Chae, Clouston, Hatzenbuehler, Kramer, Cooper,
  Wilson, Stephens-Davidowitz, Gold, and Link}{Chae et~al\mbox{.}}{2015}]%
        {Chae2015blackMortality}
\bibfield{author}{\bibinfo{person}{David~H Chae}, \bibinfo{person}{Sean
  Clouston}, \bibinfo{person}{Mark~L Hatzenbuehler}, \bibinfo{person}{Michael~R
  Kramer}, \bibinfo{person}{Hannah~LF Cooper}, \bibinfo{person}{Sacoby~M
  Wilson}, \bibinfo{person}{Seth~I Stephens-Davidowitz},
  \bibinfo{person}{Robert~S Gold}, {and} \bibinfo{person}{Bruce~G Link}.}
  \bibinfo{year}{2015}\natexlab{}.
\newblock \showarticletitle{Association between an internet-based measure of
  area racism and black mortality}.
\newblock \bibinfo{journal}{\emph{PloS one}} \bibinfo{volume}{10},
  \bibinfo{number}{4} (\bibinfo{year}{2015}), \bibinfo{pages}{e0122963}.
\newblock
\showISSN{1932-6203}


\bibitem[\protect\citeauthoryear{Chaudhry}{Chaudhry}{2015}]%
        {Chaudhry2015}
\bibfield{author}{\bibinfo{person}{Irfan Chaudhry}.}
  \bibinfo{year}{2015}\natexlab{}.
\newblock \showarticletitle{\# Hashtagging hate: Using Twitter to track racism
  online}.
\newblock \bibinfo{journal}{\emph{First Monday}} \bibinfo{volume}{20},
  \bibinfo{number}{2} (\bibinfo{year}{2015}).
\newblock
\showISSN{1396-0466}


\bibitem[\protect\citeauthoryear{Choi, Han, Paul, and Ayala}{Choi
  et~al\mbox{.}}{2011}]%
        {choi2011strategies}
\bibfield{author}{\bibinfo{person}{Kyung-Hee Choi}, \bibinfo{person}{Chong-suk
  Han}, \bibinfo{person}{Jay Paul}, {and} \bibinfo{person}{George Ayala}.}
  \bibinfo{year}{2011}\natexlab{}.
\newblock \showarticletitle{Strategies of managing racism and homophobia among
  US ethnic and racial minority men who have sex with men}.
\newblock \bibinfo{journal}{\emph{AIDS education and prevention: official
  publication of the International Society for AIDS Education}}
  \bibinfo{volume}{23}, \bibinfo{number}{2} (\bibinfo{year}{2011}),
  \bibinfo{pages}{145}.
\newblock


\bibitem[\protect\citeauthoryear{Chunara, Bouton, Ayers, and
  Brownstein}{Chunara et~al\mbox{.}}{2013}]%
        {chunara2013assessing}
\bibfield{author}{\bibinfo{person}{Rumi Chunara}, \bibinfo{person}{Lindsay
  Bouton}, \bibinfo{person}{John~W Ayers}, {and} \bibinfo{person}{John~S
  Brownstein}.} \bibinfo{year}{2013}\natexlab{}.
\newblock \showarticletitle{Assessing the online social environment for
  surveillance of obesity prevalence}.
\newblock \bibinfo{journal}{\emph{PloS one}} \bibinfo{volume}{8},
  \bibinfo{number}{4} (\bibinfo{year}{2013}), \bibinfo{pages}{e61373}.
\newblock


\bibitem[\protect\citeauthoryear{Chunara, Wisk, and Weitzman}{Chunara
  et~al\mbox{.}}{2017}]%
        {chunara2017denominator}
\bibfield{author}{\bibinfo{person}{Rumi Chunara}, \bibinfo{person}{Lauren~E
  Wisk}, {and} \bibinfo{person}{Elissa~R Weitzman}.}
  \bibinfo{year}{2017}\natexlab{}.
\newblock \showarticletitle{Denominator issues for personally generated data in
  population health monitoring}.
\newblock \bibinfo{journal}{\emph{American journal of preventive medicine}}
  \bibinfo{volume}{52}, \bibinfo{number}{4} (\bibinfo{year}{2017}),
  \bibinfo{pages}{549--553}.
\newblock


\bibitem[\protect\citeauthoryear{Clark, Anderson, Clark, and Williams}{Clark
  et~al\mbox{.}}{1999}]%
        {clark1999racism}
\bibfield{author}{\bibinfo{person}{Rodney Clark}, \bibinfo{person}{Norman~B
  Anderson}, \bibinfo{person}{Vernessa~R Clark}, {and} \bibinfo{person}{David~R
  Williams}.} \bibinfo{year}{1999}\natexlab{}.
\newblock \showarticletitle{Racism as a stressor for African Americans: A
  biopsychosocial model.}
\newblock \bibinfo{journal}{\emph{American psychologist}} \bibinfo{volume}{54},
  \bibinfo{number}{10} (\bibinfo{year}{1999}), \bibinfo{pages}{805}.
\newblock


\bibitem[\protect\citeauthoryear{Cohen}{Cohen}{1960}]%
        {cohen1960coefficient}
\bibfield{author}{\bibinfo{person}{Jacob Cohen}.}
  \bibinfo{year}{1960}\natexlab{}.
\newblock \showarticletitle{A coefficient of agreement for nominal scales}.
\newblock \bibinfo{journal}{\emph{Educational and psychological measurement}}
  \bibinfo{volume}{20}, \bibinfo{number}{1} (\bibinfo{year}{1960}),
  \bibinfo{pages}{37--46}.
\newblock


\bibitem[\protect\citeauthoryear{Collins}{Collins}{2009}]%
        {collins2009we}
\bibfield{author}{\bibinfo{person}{Dana Collins}.}
  \bibinfo{year}{2009}\natexlab{}.
\newblock \showarticletitle{``We're There and Queer'' Homonormative Mobility
  and Lived Experience among Gay Expatriates in Manila}.
\newblock \bibinfo{journal}{\emph{Gender \& Society}} \bibinfo{volume}{23},
  \bibinfo{number}{4} (\bibinfo{year}{2009}), \bibinfo{pages}{465--493}.
\newblock


\bibitem[\protect\citeauthoryear{Costa, Bandeira, and Nardi}{Costa
  et~al\mbox{.}}{2013}]%
        {Costa2013systematic}
\bibfield{author}{\bibinfo{person}{Angelo~Brandelli Costa},
  \bibinfo{person}{Denise~Ruschel Bandeira}, {and}
  \bibinfo{person}{Henrique~Caetano Nardi}.} \bibinfo{year}{2013}\natexlab{}.
\newblock \showarticletitle{Systematic review of instruments measuring
  homophobia and related constructs}.
\newblock \bibinfo{journal}{\emph{Journal of Applied Social Psychology}}
  \bibinfo{volume}{43}, \bibinfo{number}{6} (\bibinfo{year}{2013}),
  \bibinfo{pages}{1324--1332}.
\newblock
\showISSN{1559-1816}


\bibitem[\protect\citeauthoryear{Cranshaw, Schwartz, Hong, and Sadeh}{Cranshaw
  et~al\mbox{.}}{2012}]%
        {CranshawLivehoodsProject}
\bibfield{author}{\bibinfo{person}{Justin Cranshaw}, \bibinfo{person}{Raz
  Schwartz}, \bibinfo{person}{Jason~I Hong}, {and} \bibinfo{person}{Norman
  Sadeh}.} \bibinfo{year}{2012}\natexlab{}.
\newblock \showarticletitle{The livehoods project: Utilizing social media to
  understand the dynamics of a city}. In \bibinfo{booktitle}{\emph{ICWSM}}.
  \bibinfo{pages}{58}.
\newblock


\bibitem[\protect\citeauthoryear{Cranshaw and Yano}{Cranshaw and Yano}{2010}]%
        {cranshaw2010seeing}
\bibfield{author}{\bibinfo{person}{Justin Cranshaw} {and} \bibinfo{person}{Tae
  Yano}.} \bibinfo{year}{2010}\natexlab{}.
\newblock \showarticletitle{Seeing a home away from the home: Distilling
  proto-neighborhoods from incidental data with latent topic modeling}. In
  \bibinfo{booktitle}{\emph{CSSWC Workshop at NIPS}},
  Vol.~\bibinfo{volume}{10}.
\newblock


\bibitem[\protect\citeauthoryear{Darity~Jr}{Darity~Jr}{2003}]%
        {darity2003employment}
\bibfield{author}{\bibinfo{person}{William~A Darity~Jr}.}
  \bibinfo{year}{2003}\natexlab{}.
\newblock \showarticletitle{Employment discrimination, segregation, and
  health}.
\newblock \bibinfo{journal}{\emph{American Journal of Public Health}}
  \bibinfo{volume}{93}, \bibinfo{number}{2} (\bibinfo{year}{2003}),
  \bibinfo{pages}{226--231}.
\newblock


\bibitem[\protect\citeauthoryear{Davidson, Warmsley, Macy, and Weber}{Davidson
  et~al\mbox{.}}{2017}]%
        {davidson2017automated}
\bibfield{author}{\bibinfo{person}{Thomas Davidson}, \bibinfo{person}{Dana
  Warmsley}, \bibinfo{person}{Michael Macy}, {and} \bibinfo{person}{Ingmar
  Weber}.} \bibinfo{year}{2017}\natexlab{}.
\newblock \showarticletitle{Automated hate speech detection and the problem of
  offensive language}.
\newblock \bibinfo{journal}{\emph{arXiv preprint arXiv:1703.04009}}
  (\bibinfo{year}{2017}).
\newblock


\bibitem[\protect\citeauthoryear{De~Choudhury, Jhaver, Sugar, and
  Weber}{De~Choudhury et~al\mbox{.}}{2016}]%
        {de2016social}
\bibfield{author}{\bibinfo{person}{Munmun De~Choudhury},
  \bibinfo{person}{Shagun Jhaver}, \bibinfo{person}{Benjamin Sugar}, {and}
  \bibinfo{person}{Ingmar Weber}.} \bibinfo{year}{2016}\natexlab{}.
\newblock \showarticletitle{Social Media Participation in an Activist Movement
  for Racial Equality.}. In \bibinfo{booktitle}{\emph{ICWSM}}.
  \bibinfo{pages}{92--101}.
\newblock


\bibitem[\protect\citeauthoryear{Deng and Kasabov}{Deng and Kasabov}{2000}]%
        {deng2000esom}
\bibfield{author}{\bibinfo{person}{Da Deng} {and} \bibinfo{person}{Nikola
  Kasabov}.} \bibinfo{year}{2000}\natexlab{}.
\newblock \showarticletitle{ESOM: An algorithm to evolve self-organizing maps
  from online data streams}. In \bibinfo{booktitle}{\emph{Proc. IJCNN}},
  Vol.~\bibinfo{volume}{6}. IEEE, \bibinfo{pages}{3--8}.
\newblock


\bibitem[\protect\citeauthoryear{Deng, Chuang, and Lemmens}{Deng
  et~al\mbox{.}}{2009}]%
        {Deng2009spatial}
\bibfield{author}{\bibinfo{person}{Dong-Po Deng}, \bibinfo{person}{Tyng-Ruey
  Chuang}, {and} \bibinfo{person}{Rob Lemmens}.}
  \bibinfo{year}{2009}\natexlab{}.
\newblock \showarticletitle{Conceptualization of place via spatial clustering
  and co-occurrence analysis}. In \bibinfo{booktitle}{\emph{Proc. LBSN}}.
  \bibinfo{publisher}{ACM}, \bibinfo{pages}{49--56}.
\newblock
\showISBNx{1605588601}


\bibitem[\protect\citeauthoryear{Dos~Santos and Gatti}{Dos~Santos and
  Gatti}{2014}]%
        {dos2014deep}
\bibfield{author}{\bibinfo{person}{C{\'\i}cero~Nogueira Dos~Santos} {and}
  \bibinfo{person}{Maira Gatti}.} \bibinfo{year}{2014}\natexlab{}.
\newblock \showarticletitle{Deep Convolutional Neural Networks for Sentiment
  Analysis of Short Texts.}. In \bibinfo{booktitle}{\emph{COLING}}.
  \bibinfo{pages}{69--78}.
\newblock


\bibitem[\protect\citeauthoryear{Duncan, Kawachi, Subramanian, Aldstadt, Melly,
  and Williams}{Duncan et~al\mbox{.}}{2013}]%
        {duncan2013examination}
\bibfield{author}{\bibinfo{person}{Dustin~T Duncan}, \bibinfo{person}{Ichiro
  Kawachi}, \bibinfo{person}{SV Subramanian}, \bibinfo{person}{Jared Aldstadt},
  \bibinfo{person}{Steven~J Melly}, {and} \bibinfo{person}{David~R Williams}.}
  \bibinfo{year}{2013}\natexlab{}.
\newblock \showarticletitle{Examination of how neighborhood definition
  influences measurements of youths' access to tobacco retailers: a
  methodological note on spatial misclassification}.
\newblock \bibinfo{journal}{\emph{American journal of epidemiology}}
  \bibinfo{volume}{179}, \bibinfo{number}{3} (\bibinfo{year}{2013}),
  \bibinfo{pages}{373--381}.
\newblock


\bibitem[\protect\citeauthoryear{Egan, Frye, Kurtz, Latkin, Chen, Tobin, Yang,
  and Koblin}{Egan et~al\mbox{.}}{2011}]%
        {egan2011migration}
\bibfield{author}{\bibinfo{person}{James~E Egan}, \bibinfo{person}{Victoria
  Frye}, \bibinfo{person}{Steven~P Kurtz}, \bibinfo{person}{Carl Latkin},
  \bibinfo{person}{Minxing Chen}, \bibinfo{person}{Karin Tobin},
  \bibinfo{person}{Cui Yang}, {and} \bibinfo{person}{Beryl~A Koblin}.}
  \bibinfo{year}{2011}\natexlab{}.
\newblock \showarticletitle{Migration, neighborhoods, and networks: approaches
  to understanding how urban environmental conditions affect syndemic adverse
  health outcomes among gay, bisexual and other men who have sex with men}.
\newblock \bibinfo{journal}{\emph{AIDS and Behavior}} \bibinfo{volume}{15},
  \bibinfo{number}{1} (\bibinfo{year}{2011}), \bibinfo{pages}{35--50}.
\newblock


\bibitem[\protect\citeauthoryear{Finin, Murnane, Karandikar, Keller, Martineau,
  and Dredze}{Finin et~al\mbox{.}}{2010}]%
        {finin2010annotating}
\bibfield{author}{\bibinfo{person}{Tim Finin}, \bibinfo{person}{Will Murnane},
  \bibinfo{person}{Anand Karandikar}, \bibinfo{person}{Nicholas Keller},
  \bibinfo{person}{Justin Martineau}, {and} \bibinfo{person}{Mark Dredze}.}
  \bibinfo{year}{2010}\natexlab{}.
\newblock \showarticletitle{Annotating named entities in Twitter data with
  crowdsourcing}. In \bibinfo{booktitle}{\emph{Proceedings of the NAACL HLT
  2010 Workshop on Creating Speech and Language Data with Amazon's Mechanical
  Turk}}. Association for Computational Linguistics, \bibinfo{pages}{80--88}.
\newblock


\bibitem[\protect\citeauthoryear{Frias-Martinez, Soto, Hohwald, and
  Frias-Martinez}{Frias-Martinez et~al\mbox{.}}{2012}]%
        {frias2012characterizing}
\bibfield{author}{\bibinfo{person}{Vanessa Frias-Martinez},
  \bibinfo{person}{Victor Soto}, \bibinfo{person}{Heath Hohwald}, {and}
  \bibinfo{person}{Enrique Frias-Martinez}.} \bibinfo{year}{2012}\natexlab{}.
\newblock \showarticletitle{Characterizing urban landscapes using geolocated
  tweets}. In \bibinfo{booktitle}{\emph{2012 PASSAT and SocialCom}}. IEEE,
  \bibinfo{pages}{239--248}.
\newblock


\bibitem[\protect\citeauthoryear{Frye, Koblin, Chin, Beard, Blaney, Halkitis,
  Vlahov, and Galea}{Frye et~al\mbox{.}}{2010}]%
        {frye2010neighborhood}
\bibfield{author}{\bibinfo{person}{Victoria Frye}, \bibinfo{person}{Beryl
  Koblin}, \bibinfo{person}{John Chin}, \bibinfo{person}{John Beard},
  \bibinfo{person}{Shannon Blaney}, \bibinfo{person}{Perry Halkitis},
  \bibinfo{person}{David Vlahov}, {and} \bibinfo{person}{Sandro Galea}.}
  \bibinfo{year}{2010}\natexlab{}.
\newblock \showarticletitle{Neighborhood-level correlates of consistent condom
  use among men who have sex with men: a multi-level analysis}.
\newblock \bibinfo{journal}{\emph{AIDS and Behavior}} \bibinfo{volume}{14},
  \bibinfo{number}{4} (\bibinfo{year}{2010}), \bibinfo{pages}{974--985}.
\newblock


\bibitem[\protect\citeauthoryear{Galster}{Galster}{2008}]%
        {galster2008quantifying}
\bibfield{author}{\bibinfo{person}{George~C Galster}.}
  \bibinfo{year}{2008}\natexlab{}.
\newblock \showarticletitle{Quantifying the effect of neighbourhood on
  individuals: Challenges, alternative approaches, and promising directions}.
\newblock \bibinfo{journal}{\emph{Schmollers jahrbuch}} \bibinfo{volume}{128},
  \bibinfo{number}{1} (\bibinfo{year}{2008}), \bibinfo{pages}{7--48}.
\newblock
\showISSN{1439-121X}


\bibitem[\protect\citeauthoryear{Giraudel and Lek}{Giraudel and Lek}{2001}]%
        {giraudel2001comparison}
\bibfield{author}{\bibinfo{person}{JL Giraudel} {and} \bibinfo{person}{S Lek}.}
  \bibinfo{year}{2001}\natexlab{}.
\newblock \showarticletitle{A comparison of self-organizing map algorithm and
  some conventional statistical methods for ecological community ordination}.
\newblock \bibinfo{journal}{\emph{Ecological Modelling}} \bibinfo{volume}{146},
  \bibinfo{number}{1} (\bibinfo{year}{2001}), \bibinfo{pages}{329--339}.
\newblock


\bibitem[\protect\citeauthoryear{Goldberg and Levy}{Goldberg and Levy}{2014}]%
        {goldberg2014word2vec}
\bibfield{author}{\bibinfo{person}{Yoav Goldberg} {and} \bibinfo{person}{Omer
  Levy}.} \bibinfo{year}{2014}\natexlab{}.
\newblock \showarticletitle{word2vec Explained: deriving Mikolov et al.'s
  negative-sampling word-embedding method}.
\newblock \bibinfo{journal}{\emph{arXiv preprint arXiv:1402.3722}}
  (\bibinfo{year}{2014}).
\newblock


\bibitem[\protect\citeauthoryear{Green and Sheppard}{Green and
  Sheppard}{2013}]%
        {greenTweitterAuthor}
\bibfield{author}{\bibinfo{person}{Rachel Green} {and} \bibinfo{person}{John
  Sheppard}.} \bibinfo{year}{2013}\natexlab{}.
\newblock \showarticletitle{Comparing Frequency-and Style-Based Features for
  Twitter Author Identification}. In \bibinfo{booktitle}{\emph{FLAIRS
  Conference}}.
\newblock


\bibitem[\protect\citeauthoryear{Guo, Gahegan, MacEachren, and Zhou}{Guo
  et~al\mbox{.}}{2005}]%
        {guo2005multivariate}
\bibfield{author}{\bibinfo{person}{Diansheng Guo}, \bibinfo{person}{Mark
  Gahegan}, \bibinfo{person}{Alan~M MacEachren}, {and} \bibinfo{person}{Biliang
  Zhou}.} \bibinfo{year}{2005}\natexlab{}.
\newblock \showarticletitle{Multivariate analysis and geovisualization with an
  integrated geographic knowledge discovery approach}.
\newblock \bibinfo{journal}{\emph{Cartography and Geographic Information
  Science}} \bibinfo{volume}{32}, \bibinfo{number}{2} (\bibinfo{year}{2005}),
  \bibinfo{pages}{113--132}.
\newblock


\bibitem[\protect\citeauthoryear{Halkitis and Figueroa}{Halkitis and
  Figueroa}{2013}]%
        {halkitis2013sociodemographic}
\bibfield{author}{\bibinfo{person}{Perry~N Halkitis} {and}
  \bibinfo{person}{Rafael~Perez Figueroa}.} \bibinfo{year}{2013}\natexlab{}.
\newblock \showarticletitle{Sociodemographic characteristics explain
  differences in unprotected sexual behavior among young HIV-negative gay,
  bisexual, and other YMSM in New York City}.
\newblock \bibinfo{journal}{\emph{AIDS patient care and STDs}}
  \bibinfo{volume}{27}, \bibinfo{number}{3} (\bibinfo{year}{2013}),
  \bibinfo{pages}{181--190}.
\newblock


\bibitem[\protect\citeauthoryear{Hollenstein and Purves}{Hollenstein and
  Purves}{2010}]%
        {hollenstein2010exploring}
\bibfield{author}{\bibinfo{person}{Livia Hollenstein} {and}
  \bibinfo{person}{Ross Purves}.} \bibinfo{year}{2010}\natexlab{}.
\newblock \showarticletitle{Exploring place through user-generated content:
  Using Flickr tags to describe city cores}.
\newblock \bibinfo{journal}{\emph{Journal of Spatial Information Science}}
  \bibinfo{volume}{2010}, \bibinfo{number}{1} (\bibinfo{year}{2010}),
  \bibinfo{pages}{21--48}.
\newblock


\bibitem[\protect\citeauthoryear{Huang, Elghafari, Relia, and Chunara}{Huang
  et~al\mbox{.}}{2017}]%
        {huang2017high}
\bibfield{author}{\bibinfo{person}{Tom Huang}, \bibinfo{person}{Anas
  Elghafari}, \bibinfo{person}{Kunal Relia}, {and} \bibinfo{person}{Rumi
  Chunara}.} \bibinfo{year}{2017}\natexlab{}.
\newblock \showarticletitle{High-resolution temporal representations of alcohol
  and tobacco behaviors from social media data}.
\newblock \bibinfo{journal}{\emph{Proceedings of the ACM on human-computer
  interaction}} \bibinfo{volume}{1}, \bibinfo{number}{CSCW}
  (\bibinfo{year}{2017}).
\newblock


\bibitem[\protect\citeauthoryear{Intagorn and Lerman}{Intagorn and
  Lerman}{2011}]%
        {intagorn2011learning}
\bibfield{author}{\bibinfo{person}{Suradej Intagorn} {and}
  \bibinfo{person}{Kristina Lerman}.} \bibinfo{year}{2011}\natexlab{}.
\newblock \showarticletitle{Learning boundaries of vague places from noisy
  annotations}. In \bibinfo{booktitle}{\emph{Proc. ACM SIGSPATIAL}}. ACM,
  \bibinfo{pages}{425--428}.
\newblock


\bibitem[\protect\citeauthoryear{Jee-Lyn~Garc{\'\i}a and
  Sharif}{Jee-Lyn~Garc{\'\i}a and Sharif}{2015}]%
        {jee2015black}
\bibfield{author}{\bibinfo{person}{Jennifer Jee-Lyn~Garc{\'\i}a} {and}
  \bibinfo{person}{Mienah~Zulfacar Sharif}.} \bibinfo{year}{2015}\natexlab{}.
\newblock \showarticletitle{Black lives matter: a commentary on racism and
  public health}.
\newblock \bibinfo{journal}{\emph{American journal of public health}}
  \bibinfo{volume}{105}, \bibinfo{number}{8} (\bibinfo{year}{2015}),
  \bibinfo{pages}{e27--e30}.
\newblock


\bibitem[\protect\citeauthoryear{Joachims}{Joachims}{1998}]%
        {JoachimsTextCategorization}
\bibfield{author}{\bibinfo{person}{Thorsten Joachims}.}
  \bibinfo{year}{1998}\natexlab{}.
\newblock \showarticletitle{Text categorization with support vector machines:
  Learning with many relevant features}. In \bibinfo{booktitle}{\emph{European
  conference on machine learning}}. \bibinfo{publisher}{Springer},
  \bibinfo{pages}{137--142}.
\newblock


\bibitem[\protect\citeauthoryear{Johnson, Sengupta, Sch{\"o}ning, and
  Hecht}{Johnson et~al\mbox{.}}{2016}]%
        {johnson2016geography}
\bibfield{author}{\bibinfo{person}{Isaac~L Johnson}, \bibinfo{person}{Subhasree
  Sengupta}, \bibinfo{person}{Johannes Sch{\"o}ning}, {and}
  \bibinfo{person}{Brent Hecht}.} \bibinfo{year}{2016}\natexlab{}.
\newblock \showarticletitle{The geography and importance of localness in
  geotagged social media}. In \bibinfo{booktitle}{\emph{Proc. CHI}}. ACM,
  \bibinfo{pages}{515--526}.
\newblock


\bibitem[\protect\citeauthoryear{Jones and Pebley}{Jones and Pebley}{2014}]%
        {Jones2014redefiningNbhd}
\bibfield{author}{\bibinfo{person}{Malia Jones} {and} \bibinfo{person}{Anne~R
  Pebley}.} \bibinfo{year}{2014}\natexlab{}.
\newblock \showarticletitle{Redefining neighborhoods using common destinations:
  Social characteristics of activity spaces and home census tracts compared}.
\newblock \bibinfo{journal}{\emph{Demography}} \bibinfo{volume}{51},
  \bibinfo{number}{3} (\bibinfo{year}{2014}), \bibinfo{pages}{727--752}.
\newblock
\showISSN{0070-3370}


\bibitem[\protect\citeauthoryear{Kitani, Hernandez, Giraldi, and Thomaz}{Kitani
  et~al\mbox{.}}{2011}]%
        {kitani2011exploring}
\bibfield{author}{\bibinfo{person}{Edson~C Kitani}, \bibinfo{person}{Emilio~M
  Hernandez}, \bibinfo{person}{Gilson~A Giraldi}, {and}
  \bibinfo{person}{Carlos~E Thomaz}.} \bibinfo{year}{2011}\natexlab{}.
\newblock \showarticletitle{Exploring and Understanding the High Dimensional
  and Sparse Image Face Space: a Self-Organized Manifold Mapping}.
\newblock In \bibinfo{booktitle}{\emph{New Approaches to Characterization and
  Recognition of Faces}}. \bibinfo{publisher}{InTech}.
\newblock


\bibitem[\protect\citeauthoryear{Kiviluoto}{Kiviluoto}{1996}]%
        {kiviluoto1996topology}
\bibfield{author}{\bibinfo{person}{Kimmo Kiviluoto}.}
  \bibinfo{year}{1996}\natexlab{}.
\newblock \showarticletitle{Topology preservation in self-organizing maps}. In
  \bibinfo{booktitle}{\emph{IEEE Neural Networks}}, Vol.~\bibinfo{volume}{1}.
  IEEE, \bibinfo{pages}{294--299}.
\newblock


\bibitem[\protect\citeauthoryear{Koblin, Egan, Rundle, Quinn, Tieu, Cerd{\'a},
  Ompad, Greene, Hoover, and Frye}{Koblin et~al\mbox{.}}{2013}]%
        {koblin2013methods}
\bibfield{author}{\bibinfo{person}{Beryl~A Koblin}, \bibinfo{person}{James~E
  Egan}, \bibinfo{person}{Andrew Rundle}, \bibinfo{person}{James Quinn},
  \bibinfo{person}{Hong-Van Tieu}, \bibinfo{person}{Magdalena Cerd{\'a}},
  \bibinfo{person}{Danielle~C Ompad}, \bibinfo{person}{Emily Greene},
  \bibinfo{person}{Donald~R Hoover}, {and} \bibinfo{person}{Victoria Frye}.}
  \bibinfo{year}{2013}\natexlab{}.
\newblock \showarticletitle{Methods to measure the impact of home, social, and
  sexual neighborhoods of urban gay, bisexual, and other men who have sex with
  men}.
\newblock \bibinfo{journal}{\emph{PloS one}} \bibinfo{volume}{8},
  \bibinfo{number}{10} (\bibinfo{year}{2013}), \bibinfo{pages}{e75878}.
\newblock


\bibitem[\protect\citeauthoryear{Kohonen}{Kohonen}{1990}]%
        {Kohonen1990SOM}
\bibfield{author}{\bibinfo{person}{Teuvo Kohonen}.}
  \bibinfo{year}{1990}\natexlab{}.
\newblock \showarticletitle{The self-organizing map}.
\newblock \bibinfo{journal}{\emph{Proc. IEEE}} \bibinfo{volume}{78},
  \bibinfo{number}{9} (\bibinfo{year}{1990}), \bibinfo{pages}{1464--1480}.
\newblock
\showISSN{0018-9219}


\bibitem[\protect\citeauthoryear{Kreuter, Presser, and Tourangeau}{Kreuter
  et~al\mbox{.}}{2008}]%
        {Kreuter2008desirability}
\bibfield{author}{\bibinfo{person}{Frauke Kreuter}, \bibinfo{person}{Stanley
  Presser}, {and} \bibinfo{person}{Roger Tourangeau}.}
  \bibinfo{year}{2008}\natexlab{}.
\newblock \showarticletitle{Social desirability bias in CATI, IVR, and Web
  surveys the effects of mode and question sensitivity}.
\newblock \bibinfo{journal}{\emph{Public Opinion Quarterly}}
  \bibinfo{volume}{72}, \bibinfo{number}{5} (\bibinfo{year}{2008}),
  \bibinfo{pages}{847--865}.
\newblock
\showISSN{0033-362X}


\bibitem[\protect\citeauthoryear{Le~Falher, Gionis, and Mathioudakis}{Le~Falher
  et~al\mbox{.}}{2015}]%
        {LeFalherSohoRome}
\bibfield{author}{\bibinfo{person}{G{\'e}raud Le~Falher},
  \bibinfo{person}{Aristides Gionis}, {and} \bibinfo{person}{Michael
  Mathioudakis}.} \bibinfo{year}{2015}\natexlab{}.
\newblock \showarticletitle{Where is the Soho of Rome? Measures and algorithms
  for finding similar neighborhoods in cities}. In
  \bibinfo{booktitle}{\emph{ICWSM}}.
\newblock


\bibitem[\protect\citeauthoryear{Lee, Wakamiya, and Sumiya}{Lee
  et~al\mbox{.}}{2011}]%
        {lee2011discovery}
\bibfield{author}{\bibinfo{person}{Ryong Lee}, \bibinfo{person}{Shoko
  Wakamiya}, {and} \bibinfo{person}{Kazutoshi Sumiya}.}
  \bibinfo{year}{2011}\natexlab{}.
\newblock \showarticletitle{Discovery of unusual regional social activities
  using geo-tagged microblogs}.
\newblock \bibinfo{journal}{\emph{World Wide Web}} \bibinfo{volume}{14},
  \bibinfo{number}{4} (\bibinfo{year}{2011}), \bibinfo{pages}{321--349}.
\newblock


\bibitem[\protect\citeauthoryear{Liou and Yang}{Liou and Yang}{1996}]%
        {liou1996handprinted}
\bibfield{author}{\bibinfo{person}{Cheng-Yuan Liou} {and}
  \bibinfo{person}{Hsin-Chang Yang}.} \bibinfo{year}{1996}\natexlab{}.
\newblock \showarticletitle{Handprinted character recognition based on spatial
  topology distance measurement}.
\newblock \bibinfo{journal}{\emph{IEEE TPAMI}} \bibinfo{volume}{18},
  \bibinfo{number}{9} (\bibinfo{year}{1996}), \bibinfo{pages}{941--945}.
\newblock


\bibitem[\protect\citeauthoryear{Liu, Weitzman, and Chunara}{Liu
  et~al\mbox{.}}{2017}]%
        {liu2017assessing}
\bibfield{author}{\bibinfo{person}{Jason Liu}, \bibinfo{person}{Elissa~R
  Weitzman}, {and} \bibinfo{person}{Rumi Chunara}.}
  \bibinfo{year}{2017}\natexlab{}.
\newblock \showarticletitle{Assessing Behavioral Stages From Social Media
  Data}. In \bibinfo{booktitle}{\emph{Proc. CSCW}},
  Vol.~\bibinfo{volume}{2017}. NIH Public Access, \bibinfo{pages}{1320}.
\newblock


\bibitem[\protect\citeauthoryear{Liu, Deng, Shi, and Wang}{Liu
  et~al\mbox{.}}{2012}]%
        {liu2012density}
\bibfield{author}{\bibinfo{person}{Qiliang Liu}, \bibinfo{person}{Min Deng},
  \bibinfo{person}{Yan Shi}, {and} \bibinfo{person}{Jiaqiu Wang}.}
  \bibinfo{year}{2012}\natexlab{}.
\newblock \showarticletitle{A density-based spatial clustering algorithm
  considering both spatial proximity and attribute similarity}.
\newblock \bibinfo{journal}{\emph{Computers \& Geosciences}}
  \bibinfo{volume}{46} (\bibinfo{year}{2012}), \bibinfo{pages}{296--309}.
\newblock


\bibitem[\protect\citeauthoryear{MacKenzie and Podsakoff}{MacKenzie and
  Podsakoff}{2012}]%
        {mackenzie2012common}
\bibfield{author}{\bibinfo{person}{Scott~B MacKenzie} {and}
  \bibinfo{person}{Philip~M Podsakoff}.} \bibinfo{year}{2012}\natexlab{}.
\newblock \showarticletitle{Common method bias in marketing: causes,
  mechanisms, and procedural remedies}.
\newblock \bibinfo{journal}{\emph{Journal of Retailing}} \bibinfo{volume}{88},
  \bibinfo{number}{4} (\bibinfo{year}{2012}), \bibinfo{pages}{542--555}.
\newblock


\bibitem[\protect\citeauthoryear{Nagar, Yuan, Freifeld, Santillana, Nojima,
  Chunara, and Brownstein}{Nagar et~al\mbox{.}}{2014}]%
        {nagar2014case}
\bibfield{author}{\bibinfo{person}{Ruchit Nagar}, \bibinfo{person}{Qingyu
  Yuan}, \bibinfo{person}{Clark~C Freifeld}, \bibinfo{person}{Mauricio
  Santillana}, \bibinfo{person}{Aaron Nojima}, \bibinfo{person}{Rumi Chunara},
  {and} \bibinfo{person}{John~S Brownstein}.} \bibinfo{year}{2014}\natexlab{}.
\newblock \showarticletitle{A case study of the New York City 2012-2013
  influenza season with daily geocoded Twitter data from temporal and
  spatiotemporal perspectives}.
\newblock \bibinfo{journal}{\emph{Journal of medical Internet research}}
  \bibinfo{volume}{16}, \bibinfo{number}{10} (\bibinfo{year}{2014}).
\newblock


\bibitem[\protect\citeauthoryear{Noulas, Scellato, Mascolo, and Pontil}{Noulas
  et~al\mbox{.}}{2011}]%
        {noulas2011empirical}
\bibfield{author}{\bibinfo{person}{Anastasios Noulas},
  \bibinfo{person}{Salvatore Scellato}, \bibinfo{person}{Cecilia Mascolo},
  {and} \bibinfo{person}{Massimiliano Pontil}.}
  \bibinfo{year}{2011}\natexlab{}.
\newblock \showarticletitle{An empirical study of geographic user activity
  patterns in foursquare.}
\newblock \bibinfo{journal}{\emph{ICwSM}}  \bibinfo{volume}{11}
  (\bibinfo{year}{2011}), \bibinfo{pages}{70--573}.
\newblock


\bibitem[\protect\citeauthoryear{Preo{\c{t}}iuc-Pietro, Cranshaw, and
  Yano}{Preo{\c{t}}iuc-Pietro et~al\mbox{.}}{2013}]%
        {preoctiuc2013exploring}
\bibfield{author}{\bibinfo{person}{Daniel Preo{\c{t}}iuc-Pietro},
  \bibinfo{person}{Justin Cranshaw}, {and} \bibinfo{person}{Tae Yano}.}
  \bibinfo{year}{2013}\natexlab{}.
\newblock \showarticletitle{Exploring venue-based city-to-city similarity
  measures}. In \bibinfo{booktitle}{\emph{Proceedings of the 2nd ACM SIGKDD
  International Workshop on Urban Computing}}. ACM, \bibinfo{pages}{16}.
\newblock


\bibitem[\protect\citeauthoryear{Rand}{Rand}{1971}]%
        {Rand1971clusteringEval}
\bibfield{author}{\bibinfo{person}{William~M Rand}.}
  \bibinfo{year}{1971}\natexlab{}.
\newblock \showarticletitle{Objective criteria for the evaluation of clustering
  methods}.
\newblock \bibinfo{journal}{\emph{Journal of the American Statistical
  association}} \bibinfo{volume}{66}, \bibinfo{number}{336}
  (\bibinfo{year}{1971}), \bibinfo{pages}{846--850}.
\newblock
\showISSN{0162-1459}


\bibitem[\protect\citeauthoryear{Rattenbury and Naaman}{Rattenbury and
  Naaman}{2009}]%
        {Rattenbury2009placeFlickr}
\bibfield{author}{\bibinfo{person}{Tye Rattenbury} {and} \bibinfo{person}{Mor
  Naaman}.} \bibinfo{year}{2009}\natexlab{}.
\newblock \showarticletitle{Methods for extracting place semantics from Flickr
  tags}.
\newblock \bibinfo{journal}{\emph{ACM Transactions on the Web (TWEB)}}
  \bibinfo{volume}{3}, \bibinfo{number}{1} (\bibinfo{year}{2009}),
  \bibinfo{pages}{1}.
\newblock
\showISSN{1559-1131}


\bibitem[\protect\citeauthoryear{Silva, Mondal, Correa, Benevenuto, and
  Weber}{Silva et~al\mbox{.}}{2016}]%
        {silva2016analyzing}
\bibfield{author}{\bibinfo{person}{Leandro~Ara{\'u}jo Silva},
  \bibinfo{person}{Mainack Mondal}, \bibinfo{person}{Denzil Correa},
  \bibinfo{person}{Fabr{\'\i}cio Benevenuto}, {and} \bibinfo{person}{Ingmar
  Weber}.} \bibinfo{year}{2016}\natexlab{}.
\newblock \showarticletitle{Analyzing the Targets of Hate in Online Social
  Media.}. In \bibinfo{booktitle}{\emph{ICWSM}}. \bibinfo{pages}{687--690}.
\newblock


\bibitem[\protect\citeauthoryear{Srivastava, Pande, and Ranu}{Srivastava
  et~al\mbox{.}}{2015}]%
        {srivastava2015geo}
\bibfield{author}{\bibinfo{person}{Shivam Srivastava},
  \bibinfo{person}{Shiladitya Pande}, {and} \bibinfo{person}{Sayan Ranu}.}
  \bibinfo{year}{2015}\natexlab{}.
\newblock \showarticletitle{Geo-social clustering of places from check-in
  data}. In \bibinfo{booktitle}{\emph{Data Mining (ICDM), 2015 IEEE
  International Conference on}}. IEEE, \bibinfo{pages}{985--990}.
\newblock


\bibitem[\protect\citeauthoryear{Stephens-Davidowitz}{Stephens-Davidowitz}{2014}]%
        {stephens2014cost}
\bibfield{author}{\bibinfo{person}{Seth Stephens-Davidowitz}.}
  \bibinfo{year}{2014}\natexlab{}.
\newblock \showarticletitle{The cost of racial animus on a black candidate:
  Evidence using Google search data}.
\newblock \bibinfo{journal}{\emph{Journal of Public Economics}}
  \bibinfo{volume}{118} (\bibinfo{year}{2014}), \bibinfo{pages}{26--40}.
\newblock


\bibitem[\protect\citeauthoryear{Tang, Wei, Yang, Zhou, Liu, and Qin}{Tang
  et~al\mbox{.}}{2014}]%
        {tang2014learning}
\bibfield{author}{\bibinfo{person}{Duyu Tang}, \bibinfo{person}{Furu Wei},
  \bibinfo{person}{Nan Yang}, \bibinfo{person}{Ming Zhou},
  \bibinfo{person}{Ting Liu}, {and} \bibinfo{person}{Bing Qin}.}
  \bibinfo{year}{2014}\natexlab{}.
\newblock \showarticletitle{Learning Sentiment-Specific Word Embedding for
  Twitter Sentiment Classification.}. In \bibinfo{booktitle}{\emph{ACL (1)}}.
  \bibinfo{pages}{1555--1565}.
\newblock


\bibitem[\protect\citeauthoryear{Tobin, Cutchin, Latkin, and Takahashi}{Tobin
  et~al\mbox{.}}{2013}]%
        {tobin2013social}
\bibfield{author}{\bibinfo{person}{KE Tobin}, \bibinfo{person}{M Cutchin},
  \bibinfo{person}{CA Latkin}, {and} \bibinfo{person}{LM Takahashi}.}
  \bibinfo{year}{2013}\natexlab{}.
\newblock \showarticletitle{Social geographies of African American men who have
  sex with men (MSM): A qualitative exploration of the social, spatial and
  temporal context of HIV risk in Baltimore, Maryland}.
\newblock \bibinfo{journal}{\emph{Health \& place}}  \bibinfo{volume}{22}
  (\bibinfo{year}{2013}), \bibinfo{pages}{1--6}.
\newblock


\bibitem[\protect\citeauthoryear{Tourangeau, Rasinski, and Bradburn}{Tourangeau
  et~al\mbox{.}}{1991}]%
        {tourangeau1991measuring}
\bibfield{author}{\bibinfo{person}{Roger Tourangeau},
  \bibinfo{person}{Kenneth~A Rasinski}, {and} \bibinfo{person}{Norman
  Bradburn}.} \bibinfo{year}{1991}\natexlab{}.
\newblock \showarticletitle{Measuring happiness in surveys: A test of the
  subtraction hypothesis}.
\newblock \bibinfo{journal}{\emph{Public Opinion Quarterly}}
  \bibinfo{volume}{55}, \bibinfo{number}{2} (\bibinfo{year}{1991}),
  \bibinfo{pages}{255--266}.
\newblock


\bibitem[\protect\citeauthoryear{Waseem and Hovy}{Waseem and Hovy}{2016}]%
        {waseem2016hateful}
\bibfield{author}{\bibinfo{person}{Zeerak Waseem} {and} \bibinfo{person}{Dirk
  Hovy}.} \bibinfo{year}{2016}\natexlab{}.
\newblock \showarticletitle{Hateful Symbols or Hateful People? Predictive
  Features for Hate Speech Detection on Twitter.}. In
  \bibinfo{booktitle}{\emph{SRW@ HLT-NAACL}}. \bibinfo{pages}{88--93}.
\newblock


\bibitem[\protect\citeauthoryear{White}{White}{1983}]%
        {White1983spatial}
\bibfield{author}{\bibinfo{person}{Michael~J White}.}
  \bibinfo{year}{1983}\natexlab{}.
\newblock \showarticletitle{The measurement of spatial segregation}.
\newblock \bibinfo{journal}{\emph{American journal of sociology}}
  (\bibinfo{year}{1983}), \bibinfo{pages}{1008--1018}.
\newblock
\showISSN{0002-9602}


\bibitem[\protect\citeauthoryear{Williams and Burnap}{Williams and
  Burnap}{2015}]%
        {williams2015cyberhate}
\bibfield{author}{\bibinfo{person}{Matthew~L Williams} {and}
  \bibinfo{person}{Pete Burnap}.} \bibinfo{year}{2015}\natexlab{}.
\newblock \showarticletitle{Cyberhate on social media in the aftermath of
  Woolwich: A case study in computational criminology and big data}.
\newblock \bibinfo{journal}{\emph{British Journal of Criminology}}
  \bibinfo{volume}{56}, \bibinfo{number}{2} (\bibinfo{year}{2015}),
  \bibinfo{pages}{211--238}.
\newblock


\bibitem[\protect\citeauthoryear{Yin}{Yin}{2002}]%
        {yin2002data}
\bibfield{author}{\bibinfo{person}{Hujun Yin}.}
  \bibinfo{year}{2002}\natexlab{}.
\newblock \showarticletitle{Data visualisation and manifold mapping using the
  ViSOM}.
\newblock \bibinfo{journal}{\emph{Neural Networks}} \bibinfo{volume}{15},
  \bibinfo{number}{8} (\bibinfo{year}{2002}), \bibinfo{pages}{1005--1016}.
\newblock


\end{thebibliography}

\end{document}